\documentclass[12pt,preprint]{aastex}
\usepackage{hyperref}
\usepackage{breakurl}

\def\M20{$M_{20}$}
\def\micron{$\mu$m}

\shorttitle{}
\shortauthors{Farrah et. al.}

\begin{document}

\title{The role of the most luminous, obscured AGN in galaxy assembly at $\rm{z} \sim 2$}

\author{Duncan Farrah\altaffilmark{1}}
\author{Sara Petty\altaffilmark{2}}
\author{Brian Connolly\altaffilmark{3}}
\author{Andrew Blain\altaffilmark{4}}
\author{Andreas Efstathiou\altaffilmark{5}}
\author{Mark Lacy\altaffilmark{6}}
\author{Daniel Stern\altaffilmark{7}}
\author{Sean Lake\altaffilmark{8}}
\author{Tom Jarrett\altaffilmark{9}}
\author{Carrie Bridge\altaffilmark{7}}
\author{Peter Eisenhardt\altaffilmark{7}}
\author{Dominic Benford\altaffilmark{10}}
\author{Suzy Jones\altaffilmark{11}}
\author{Chao-Wei Tsai\altaffilmark{8}}
\author{Roberto Assef\altaffilmark{12}}
\author{Jingwen Wu\altaffilmark{13}}
\author{Leonidas Moustakas\altaffilmark{7}}

\altaffiltext{1}{Department of Physics, Virginia Tech, Blacksburg, VA  24061, USA} 
\altaffiltext{2}{Green Science Policy Institute, Berkeley, CA 94709, USA}
\altaffiltext{3}{Cincinnati Children's Hospital Medical Center, 3333 Burnet Ave, Cincinnati, OH 45229}
\altaffiltext{4}{Department of Physics and Astronomy, University of Leicester, Leicester LE1 7RH, UK}
\altaffiltext{5}{School of Sciences, European University Cyprus, Diogenes Street, Engomi, 1516 Nicosia, Cyprus}
\altaffiltext{6}{National Radio Astronomy Observatory, 520 Edgemont Road, Charlottesville, VA 22903, USA} 
\altaffiltext{7}{Jet Propulsion Laboratory, California Institute of Technology, 4800 Oak Grove Drive, Pasadena, CA 91109, USA}
\altaffiltext{8}{Physics and Astronomy Department, University of California, Los Angeles, CA 90095, USA}
\altaffiltext{9}{Department of Astronomy, University of Cape Town, 7700 Rondebosch, South Africa} 
\altaffiltext{10}{Observational Cosmology Lab., Code 665, NASA at Goddard Space Flight Center, Greenbelt, MD 20771, USA} 
\altaffiltext{11}{Department of Space, Earth, and Environment, Chalmers University of Technology, Onsala Space Observatory, SE-43992 Onsala, Sweden} 
\altaffiltext{12}{N\'ucleo de Astronom\'ia de la Facultad de Ingenier\'ia, Universidad Diego Portales, Av. Ej\'ercito Libertador 441, Santiago, Chile}
\altaffiltext{13}{National Astronomical Observatories, Chinese Academy of Sciences, 20A Datun Road, Chaoyang District, Beijing, 100012, China}

\begin{abstract}
We present HST WFC3 F160W imaging and infrared spectral energy distributions for twelve extremely luminous, obscured AGN at $1.8<z<2.7$, selected via ``Hot, Dust Obscured'' mid-infrared colors. Their infrared luminosities span $2-15\times10^{13}$\,L$_{\odot}$, making them among the most luminous objects in the Universe at $z\sim2$. In all cases the infrared emission is consistent with arising at least in most part from AGN activity. The AGN fractional luminosities are higher than those in either sub-millimeter galaxies, or AGN selected via other mid-infrared criteria. Adopting the $G$, \M20 and $A$ morphological parameters, together with traditional classification boundaries, infers that three quarters of the sample as mergers. Our sample do not, however, show any correlation between the considered morphological parameters and either infrared luminosity or AGN fractional luminosity. Moreover, their asymmetries and effective radii are distributed identically to those of massive galaxies at $z\sim2$. We conclude that our sample is not preferentially associated with mergers, though a significant merger fraction is still plausible. Instead, we propose that our sample are examples of the massive galaxy population at $z\sim2$ that harbor a briefly luminous, ``flickering'' AGN, and in which the $G$ and \M20 values have been perturbed, due to either the AGN, and/or the earliest formation stages of a bulge in an inside-out manner. Furthermore, we find that the mass assembly of the central black holes in our sample leads the mass assembly of any bulge component. Finally, we speculate that our sample represent a small fraction of the immediate antecedents of compact star-forming galaxies at $z\sim2$.
 \end{abstract}

\keywords{galaxies: starburst}

\section{Introduction}
Active Galactic Nuclei (AGN) play a fundamental role in galaxy assembly. AGN signpost the relatively brief periods in a galaxy's lifetime when the central supermassive black hole is accreting rapidly, during which the black hole likely assembled the bulk of its mass \citep{DrapBal2012,treister12}. There is also a deep connection between AGN activity and star formation rates at all redshifts. The most striking evidence for this connection is the similar cosmological evolution of AGN and star formation; the optical luminosity function of quasars plateaus between $2 < z < 3$ \citep[e.g.][]{richards06qsolf,Delvecchio2014}, while the comoving star formation rate density also peaks at $z\sim2$ \citep[e.g.][]{Connolly1997,Lanzetta2002,hop06}, with a decline towards both lower \citep[e.g.][]{Lilly1996,Dickinson2003,LeFloch05,igl07,wang13} and higher redshifts \citep[e.g.][]{PerezGonzalez2005,wall08,Wuyts2011,Bethermin2012}. Further evidence for this connection includes the $M_{bh}-\sigma$ relation \citep[e.g.][]{Magor98,Tremaine2002}, the presence of luminous, coeval starbursts and AGN in galaxies \citep{farrah03,alexander05,lonsdale2006,hernan09,spoo13}, and observed scaling relations between AGN properties and star formation rates in luminous quasars \citep{harris2016}. Finally, star formation and AGN activity may directly affect each other \citep{fabian12}, via both quenching \citep[e.g.][]{dima05,bower06,croton06,farrah12,Alatalo2015} and triggering \citep[e.g.][]{deyo89,king05,cro06,gaib12,silk13,Zubovas2013,ishi13}. 

The role of AGN during the peak epoch of galaxy assembly at $z\sim2$ can be studied by identifying active galaxies at high redshift, characterizing the power sources within them, and determining which mechanisms trigger these power sources. Doing so however faces two challenges. First, star-forming regions and AGN are often occulted by large column densities of gas and dust. Thus, a substantial fraction of their light is observed in the infrared \citep{san96,lagache05,farrah13,Casey2014_review}. Second, at high redshifts, obscured systems are seen both faintly and at coarsened spatial scales, making them difficult to identify.

\begin{deluxetable}{lcccccccccc}
\rotate
\tabletypesize{\scriptsize}
\tablecaption{The WISE-selected Hot DOG Sample, and their flux densities from HST, WISE, and {\itshape Herschel} \label{table:basic}}
\tablehead{
\colhead{Object}                             &\colhead{HST}    &\multicolumn{4}{c}{WISE}                                        &\multicolumn{2}{c}{$Herschel$-PACS} & \multicolumn{3}{c}{$Herschel$-SPIRE}         \\
\colhead{}                                   &\colhead{1.6$\mu$m}      & \colhead{3.4$\mu$m}   & \colhead{4.6$\mu$m}   & \colhead{12$\mu$m}& \colhead{22$\mu$m} &\colhead{70$\mu$m} & \colhead{160$\mu$m}      &\colhead{250$\mu$m}&\colhead{350$\mu$m}&\colhead{500$\mu$m}   \\
\colhead{}                                   &\colhead{($\mu$Jy)}&\colhead{($\mu$Jy)}&\colhead{($\mu$Jy)}&\colhead{(mJy)}&\colhead{(mJy)} &\colhead{(mJy)}&\colhead{(mJy)}&\colhead{(mJy)}&\colhead{(mJy)}&\colhead{(mJy)}
}
\startdata
 WISEA J042138.55$+$203644.6                  &  9.2$\pm$0.8      &  11.0$\pm$6.6   &  25.0$\pm$9.7   & 2.7$\pm$0.2 &  9.2$\pm$1.1 & 30$\pm$5    & 17$\pm$18   & 27$\pm$10   & 20$\pm$15   & 17$\pm$10  \\
 WISEA J051442.62$-$121724.4                  & 67.9$\pm$2.1      & 107.0$\pm$5.7   & 124.0$\pm$10.2  & 5.9$\pm$0.2 & 30.1$\pm$1.3 & 105$\pm$11  & 165$\pm$20  & 110$\pm$12  & 51$\pm$10   & 18$\pm$9   \\
 WISEA J054230.90$-$270539.8\tablenotemark{a} &  7.8$\pm$0.7      &  14.8$\pm$3.8   &  30.8$\pm$12.3  & 2.8$\pm$0.1 & 14.7$\pm$1.0 & 43$\pm$4    & 63$\pm$11   & 47$\pm$6    & 23$\pm$10   & 16$\pm$11  \\
 WISEA J060508.95$-$232434.5                  &  8.2$\pm$0.7      &  24.3$\pm$3.9   &  76.9$\pm$8.4   & 3.9$\pm$0.2 & 16.3$\pm$1.0 & 34$\pm$5    & 117$\pm$20  & 103$\pm$15  & 58$\pm$10   & 26$\pm$10  \\
 WISEA J091247.26$+$774158.2                  & 13.7$\pm$0.9      &  73.7$\pm$2.8   &  25.3$\pm$7.0   & 1.9$\pm$0.1 &  8.3$\pm$7.8 & 27$\pm$6    & 17$\pm$12   & 15$\pm$12   & 11$\pm$10   & 9$\pm$15   \\
 WISEA J120629.69$+$623224.3                  &  6.9$\pm$0.7      &  18.7$\pm$3.4   &  27.3$\pm$6.7   & 2.3$\pm$0.1 & 11.8$\pm$0.9 & 37$\pm$7    & 30$\pm$13   & 33$\pm$10   & 14$\pm$11   & 16$\pm$18  \\
 WISEA J131628.54$+$351235.6\tablenotemark{b} &  9.8$\pm$0.8      &  16.6$\pm$3.6   &  39.3$\pm$7.7   & 3.3$\pm$0.1 & 12.8$\pm$1.0 & 27$\pm$6    & 41$\pm$4    & 39$\pm$8    & 12$\pm$10   & 11$\pm$13  \\
 WISEA J171946.63$+$044635.4                  & 19.8$\pm$1.1      &  28.5$\pm$4.5   & 109.0$\pm$10.5  & 5.2$\pm$0.2 & 15.2$\pm$1.2 & 35$\pm$6    & 43$\pm$8    & 69$\pm$6    & 56$\pm$5    & 40$\pm$7   \\
 WISEA J182242.67$+$634853.4                  &  8.9$\pm$0.8      &  37.2$\pm$2.2   & 134.0$\pm$5.2   & 3.6$\pm$0.1 & 12.1$\pm$0.6 & 26$\pm$5    & 10$\pm$12   & 9$\pm$10    & 12$\pm$10   & 6$\pm$14   \\
 WISEA J183013.51$+$650420.4\tablenotemark{c} &  1.5$\pm$0.3      &  11.4$\pm$1.6   &  13.1$\pm$3.3   & 2.4$\pm$0.1 &  7.4$\pm$0.5 & 17$\pm$6    & 25$\pm$10   & 14$\pm$7    & 16$\pm$9    & 7$\pm$13   \\
 WISEA J183533.73$+$435548.7\tablenotemark{d} & 12.8$\pm$0.9      &  32.0$\pm$2.6   & 139.0$\pm$6.5   & 6.8$\pm$0.2 & 27.4$\pm$1.0 & 54$\pm$5    & 81$\pm$21   & 113$\pm$12  & 82$\pm$9    & 46$\pm$7   \\
 WISEA J233759.50$+$792654.6                  & 21.8$\pm$1.2      &  31.0$\pm$4.0   &  70.3$\pm$6.4   & 2.3$\pm$0.1 & 14.6$\pm$0.9 & 64$\pm$7    & 101$\pm$12  & 58$\pm$7    & 38$\pm$6    & 17$\pm$10  \\
\enddata
\tablecomments{We also include ground-based far-infrared photometry from \citealt{wu12,jones14}, listed below.}
\tablenotetext{a}{$<47$mJy at 350$\mu$m}
\tablenotetext{b}{$<14.2$mJy at 1100$\mu$m}
\tablenotetext{c}{$<31$mJy at 350$\mu$m}
\tablenotetext{d}{$46\pm16$mJy at 350$\mu$m, $31\pm14$mJy at 450$\mu$m, $8\pm1.5$mJy at 850$\mu$m}
\end{deluxetable}

High redshift obscured systems can be identified via several approaches, including hard X-ray flux, rest-frame optical line ratios, and infrared photometry. In the case of infrared photometry the selection is based on colors involving at least one infrared band, where the choice of bands predisposes the selection to sources with different effective dust temperatures. This includes sources with cold (up to about 40\,K) dust heated by star formation in the case of sub-mm selection, or hotter dust heated by AGN in the case of mid-infrared color selection. In the case of selections that use one optical and one infrared band and then demand an excess in the infrared band, such as the `dust obscured galaxy' (DOG) selection, the result is usually a mixture of AGN and starburst dominated systems \citep{dey08}. 

Determining the power source of obscured systems is more challenging. In the local Universe, it is possible to diagnose obscured power sources with reasonable accuracy, leading to the consensus that the majority of systems with infrared luminosities up to about $10^{12}$L$_{\odot}$ (the Luminous Infrared Galaxies, or LIRGs) are starburst dominated \citep{sti13,petty14}. At infrared luminosities exceeding $10^{12}$L$_{\odot}$ (the Ultraluminous Infrared Galaxies, or ULIRGs) there is a greater contribution from obscured AGN \citep{gen98,farrah03}. At higher redshifts, however, such diagnoses are harder. For example, sub-mm selected galaxies often contain obscured, luminous AGN, despite the predilection of their selection for star formation \citep{alexander05}. Other selections, such as the DOG selection, require additional diagnostics that use mid-infrared continuum shape to classify sources as starburst or AGN-dominated. 
 
Establishing what mechanisms trigger their infrared emission faces conceptually similar challenges. At low redshifts it is straightforward to quantify morphologies as a route to answering this question; LIRGs have diverse morphologies, but the ULIRGs are almost exclusively mergers \citep{sur98,farrah02,bridge07,haan11}. The greater diversity in LIRG morphologies may reflect the broader set of evolutionary pathways that a galaxy can take through a LIRG, rather than the more luminous ULIRG, phase \citep[e.g.][]{far09}. Outside the local Universe however there remains significant uncertainty. For ULIRGs, the merger fraction probably declines by at most a small amount between $z=0$ and $z=1$ \citep{hung14} but the behavior at $z>1$ is less clear. Some studies find that $z\gtrsim1$ infrared-luminous systems are mostly mergers \citep{ricc10,zam11,alag12,ivison12,kar12,chen15,oli16}, while others find that they are not \citep{mel09,ricc10,agui13,wik14}, instead resembling either early types \citep{swin10} or disks \citep{targ11,sch12,tac13,targ13}. This disagreement is mirrored by theoretical work; some models invoke mergers \citep{baugh05,chak08,hay11,hop10}, some invoke `secular' processes \citep{dek06,dek09,gen08,ker09,dav10,nara15}, and others use both \citep{hay13}. 

\begin{deluxetable}{lcccccccc}
\tabletypesize{\scriptsize}
\tablecaption{Redshifts, luminosities and morphological parameters \label{table:lummorph}}
\tablehead{
\colhead{Object}       &\colhead{$z$} & \colhead{$L\mathrm{_{IR}}$}    & \colhead{$f\mathrm{_{AGN}}$}&\colhead{$G$}       &\colhead{\M20}      &\colhead{$A$}    &\colhead{$n$}& \colhead{$r_{e}$} \\
\colhead{}             &\colhead{}    & \colhead{($10^{13}L_{\odot}$)} &                             &                    &                    &                 &             & \colhead{(kpc)}      
 }                                                                     
 \startdata                                                                      
WISEA 0421 & 1.83         & $2.09 \pm 0.18$              & $1.00^{+u}_{-0.13}$         & $ 0.575 \pm 0.031$ & $-1.822 \pm 0.146$ & $0.26 \pm 0.03$ & $1.20 $ & $3.34 $ \\
WISEA 0514 & 2.50         & $14.2 \pm 1.00$              & $0.91^{+0.04}_{-0.04}$      & $ 0.811 \pm 0.045$ & $-2.260 \pm 0.203$ & $0.39 \pm 0.07$ & $0.70 $ & $2.54 $ \\
WISEA 0542 & 2.53         & $6.77 \pm 0.50$              & $0.90^{+0.02}_{-0.03}$      & $ 0.775 \pm 0.037$ & $-1.520 \pm 0.205$ & $0.33 \pm 0.04$ & $1.01 $ & $1.38 $ \\
WISEA 0605 & 2.08         & $4.78 \pm 0.25$              & $0.69^{+0.06}_{-0.05}$      & $ 0.674 \pm 0.035$ & $-2.278 \pm 0.202$ & $0.16 \pm 0.02$ & $2.01 $ & $3.36 $ \\
WISEA 0912 & 2.00         & $1.87 \pm 0.30$              & $0.95^{+0.05}_{-0.10}$      & $ 0.745 \pm 0.046$ & $-2.102 \pm 0.362$ & $0.18 \pm 0.03$ & $1.27 $ & $2.43 $ \\
WISEA 1206 & 2.00         & $2.57 \pm 0.50$              & $1.00^{+u}_{-0.25}$         & $ 0.628 \pm 0.027$ & $-2.460 \pm 0.258$ & $0.17 \pm 0.03$ & $2.12 $ & $5.40 $ \\
WISEA 1316 & 1.96         & $3.03 \pm 0.30$              & $0.83^{+0.10}_{-0.10}$      & $ 0.784 \pm 0.045$ & $-1.941 \pm 0.351$ & $0.24 \pm 0.04$ & $1.50 $ & $1.72 $ \\
WISEA 1719 & 2.54         & $8.76 \pm 0.90$              & $0.99^{+0.01}_{-0.02}$      & $ 0.804 \pm 0.043$ & $-2.044 \pm 0.183$ & $0.74 \pm 0.13$ & $1.32 $ & $4.03 $ \\
WISEA 1822 & 2.07         & $3.30 \pm 0.30$              & $0.98^{+0.02}_{-0.07}$      & $ 0.663 \pm 0.037$ & $-2.073 \pm 0.177$ & $0.29 \pm 0.05$ & $0.29 $ & $6.21 $ \\
WISEA 1830 & 2.65         & $4.41 \pm 0.35$              & $0.96^{+0.04}_{-0.03}$      & $ 0.525 \pm 0.025$ & $-0.878 \pm 0.119$ & $0.24 \pm 0.04$ & $1.18 $ & $2.80 $ \\
WISEA 1835 & 2.30         & $9.62 \pm 0.50$              & $0.83^{+0.07}_{-0.10}$      & $ 0.472 \pm 0.026$ & $-1.362 \pm 0.114$ & $0.22 \pm 0.03$ & $1.33 $ & $1.81 $ \\
WISEA 2337 & 2.74         & $10.1 \pm 1.00$              & $0.90^{+0.08}_{-0.15}$      & $ 0.826 \pm 0.034$ & $-1.808 \pm 0.107$ & $0.37 \pm 0.05$ & $1.34 $ & $1.18 $ \\ 
 \enddata                            
 \tablecomments{The methods used to compute these quantities are described in \S\ref{analysis}. The uncertainties on the effective radii and S{\'e}rsic indices are approximately 15\% in all cases.}
 \end{deluxetable}

For these reasons, it is valuable to employ photometric selections that isolate different populations of high redshift infrared-luminous galaxies, and then use both diagnostics of their power source and measures of their morphology to place them in the wider context of galaxy assembly. Moreover, it is valuable to study the {\itshape most luminous} AGN in the Universe - systems with bolometric luminosities exceeding $\sim10^{47}$ ergs s$^{-1}$. Although rare, such systems probe the role of AGN in galaxy assembly at their most extreme limits, and can supply stringent tests for galaxy evolution models since they imply sustained, very high accretion rates of $\gg10$M$_{\odot}$yr$^{-1}$. 

In this paper we undertake such a study with a sample of twelve extremely infrared-luminous AGN-dominated systems at $z\sim2$, all with spectroscopic redshifts, selected with data from the Wide-field Infrared Survey Explorer (WISE, \citealt{wri10}). We then use both multi-band infrared photometry and HST imaging to elucidate their evolutionary status. We adopt Vega magnitudes, and assume a spatially flat cosmology with $\Omega_{\rm{m}} = 0.3$ and $H_{0}=70\,$km s$^{-1}$ Mpc$^{-1}$.

\begin{figure*} 
\begin{center}
\includegraphics[width=16.5cm,angle=0]{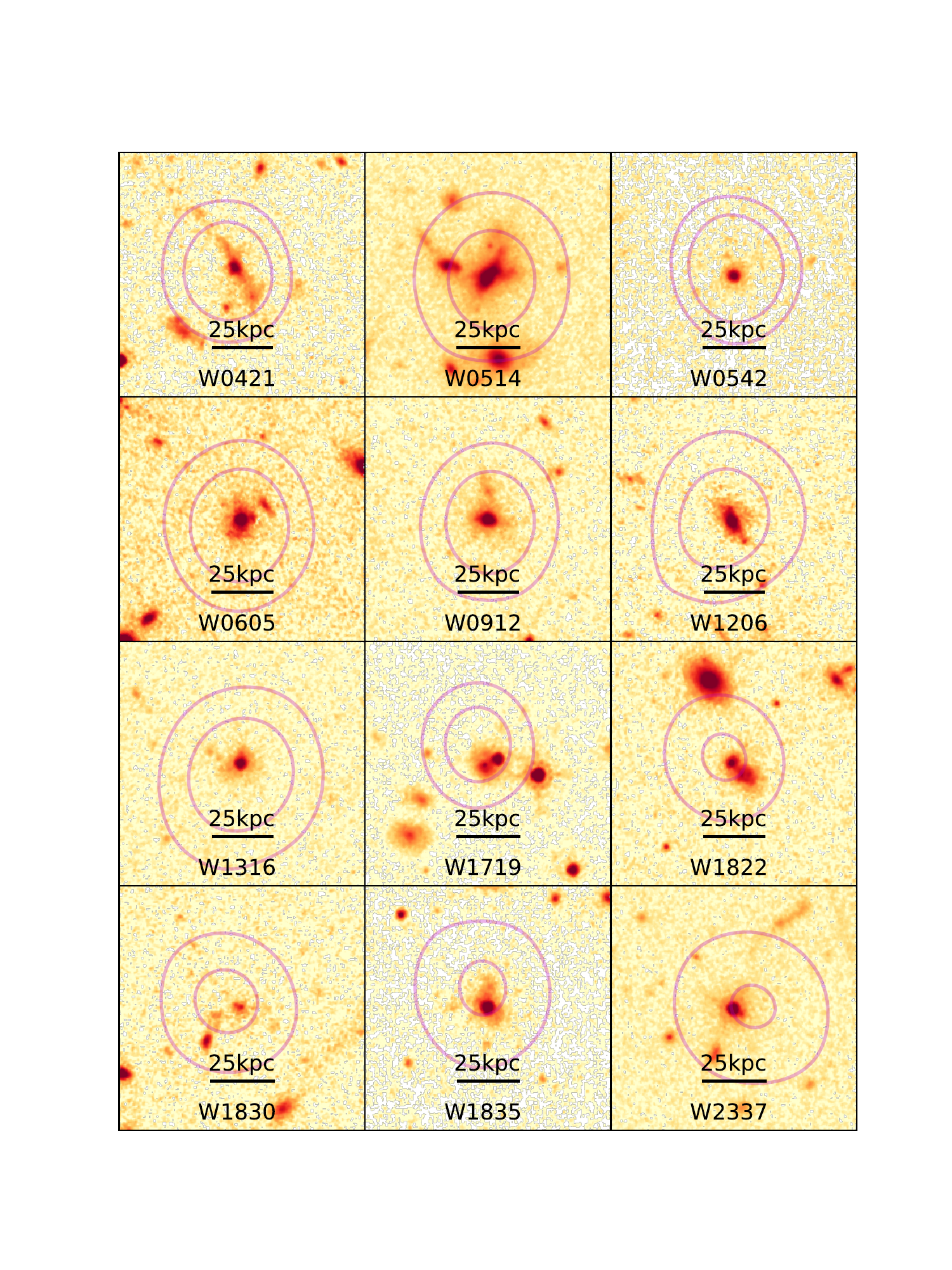} 
\vspace{-2.5cm}
\caption{The $H$-band morphologies of the twelve objects in our sample. For each object, the panel shows the WFC3 image with the W3 contours overlaid. We do not show the WISE, PACS or SPIRE images as the detections are always point sources that are well-centered on the HST sources.}
\label{fig:imagesa}
\end{center}
\end{figure*}

\section{Sample Selection}\label{sample}
Searches within the WISE color-space have isolated populations ranging from planets to galaxies \citep{cush11,griff11,eis12,stern12,wu12,ass13,Bridge2013,lon15}. For high redshift infrared-luminous galaxies, an approach that has proved successful is to search for objects that are clearly detected in both the $12\,\mu$m (W3) and $22\,\mu$m (W4) channels, but are weakly or not detected in the $3.4\,\mu$m (W1) and $4.6\,\mu$m (W2) channels \citep{eis12,Bridge2013}. The targets in this paper are selected using this approach.

The parent sample of 53 objects was selected from the AllWISE catalog \citep{cutri14}, and includes all objects with a spectroscopic redshift as of April 2013. These redshifts span $1.6<z<4.6$. Since we are selecting those objects from the parent catalog that have redshifts, there is the possibility of bias towards higher surface brightness systems with bright emission lines. We do not, however, consider this bias in our analysis as it is impossible to quantify with existing data.

The objects in the parent sample are selected in two ways. First is that of \citet{eis12}, who demand that sources be detected at $>5\sigma$ in both the W3 and W4 channels, and have a W1 flux of $<34\mathrm{\,\mu Jy}$. Second is that of \citet{Bridge2013}, who also demand that sources be detected at $>5\,\sigma$ in both the W3 and W4 channels, a non-detection in Sloan Digital Sky Survey $r'$ imaging (i.e. $r' \geq 22$), but then demand only that $W2-W3 \geq 4.8$. Both selections result in a source density of one per several square degrees \citep{ass15}. These selections lie within the Dust-Obscured Galaxy (DOG) selection: $f_{22}>0.3$mJy and $f_{22}/f_{R}>1000$ \citep{dey08}, but result in samples with higher dust temperatures, on average; the DOGs have $T_{d}\sim30-40$K \citep{pope08,mel12}, whereas our selections give $T_{d}\simeq60$K \citep{Bridge2013,jones14}. While the \citet{eis12} and \citet{Bridge2013} selections differ in detail, hereafter we treat them as identical and refer to them as ``Hot DOGS'', or $h$DOGS \citep{wu12}.

\section{Observations \label{observe}}

\subsection{HST observations}
We submitted the 53 objects in the parent sample for a Cycle 19 {\itshape Hubble Space Telescope} (HST) SNAP program (HST-GO-12585, PI Petty), of which 12 were observed (Table \ref{table:basic}). These 12 objects have spectroscopic redshifts in the range $1.8<z<2.7$. Other than a slightly lower median redshift, these 12 objects are statistically indistinguishable from the parent sample. None of their spectra show evidence for foreground lenses. 

The 12 objects were observed with the Wide Field Camera 3 (WFC3) in the F160W filter ($H$-band, hereafter). To facilitate the SNAP observations the total exposure time per object was 1500\,s, selected as it allowed each object to be observed within 48 minutes, after accounting for guide star acquisition and instrument overheads. This exposure time reaches a surface brightness limit in $H$ of $23.5$\,mag arcsec$^{-2}$. Each exposure was divided into four equal length sub-exposures, using a dither box pattern of four pointings with a 0.6\arcsec\ spacing.

All of the sample were clearly detected by WFC3. The WFC3 data were reduced using a standard Multidrizzle process. We started with the persistence-corrected output files from the {\itshape calwf3} pipeline, which performs standard tasks including bias and dark current subtraction, linearity correction, flat fielding, bad pixel masking, and removal of cosmic rays. We then removed geometric distortion from each file and combined them into a single image for each object, using the Autodrizzle task. To extract fluxes we used SExtractor \citep{bertin96} in MAG\_BEST mode, which first determines the most appropriate elliptical aperture to use, and then measures the flux inside that aperture. Finally, we corrected the photometry for Galactic absorption and converted the fluxes into magnitudes using the zeropoints provided by the WFC3 team.

\subsection{Ancillary data}
We obtained reduced {\itshape Spitzer} \citep{werner04} images at $3.6$ and $4.5\,\mu$m from the IRAC instrument \citep{faz04} for all our sample. These observations are significantly deeper than the W1 and W2 observations. For photometry, we adopted the methods in \citet{lac05}. We required $2\sigma$ levels for the detect and analysis threshold parameters, and checked each object map for false detections. 

We obtained {\itshape Herschel} \citep{pilb10} photometry for all of our sample from the Herschel Science Archive. The data were taken with the Photoconductor Array Camera and Spectrometer (PACS; \citealt{pogl10}) at both $70\,\mu$m and $160\,\mu$m, and with the Spectral and Photometeric Imaging REceiver instrument (SPIRE; \citealt{Griffin2010}) at $250\,\mu$m, $350\,\mu$m and $500\,\mu$m. The level 1 data were processed to level 2 using the {\itshape Herschel} Interactive Processing Environment (HIPE) v14.2.0. For PACS, aperture photometry was carried out using the scanmap\_pointsources\_PhotProject.py HIPE script. For SPIRE, photometry was carried out using the Sussextractor algorithm \citep{sao07,wang14}. The complete set of {\itshape Herschel} data for the $h$DOGs, including those without HST data, is presented in C-W Tsai et al, in preparation.

\section{Methods}\label{analysis}

\subsection{Morphologies}\label{submorph}
We use five parameters to quantify the morphologies of our sample; Gini ($G$), \M20, asymmetry ($A$), S{\'e}rsic index ($n$), and effective radius ($r_{e}$). 

The Gini coefficient \citep{abra03} is a measure of how concentrated the light is in an image: 

\begin{equation}
G = \frac{1}{\displaystyle |\bar{f}|N(N-1)}\sum_{i=1}^{N}(2i-N-1)|f_{i}| ,
\end{equation}

\noindent where $|\bar{f}|$, and $|f_i|$ are the absolute average flux, and $i$th pixel flux from $N$ total pixels, respectively. $G$ ranges from zero to unity, with low values for galaxies with even light distributions and high values for galaxies whose light is concentrated into a small number of bright nuclear pixels. The \M20 coefficient \citep{lotz04} is the second-order moment of the brightest 20\% of the light: 

\begin{equation}
M_{20} = \log_{10} \left( \frac{\sum_i M_i}{M_{\mathrm{tot}}} \right) ,
\end{equation}

\noindent where: 

\begin{equation}
M_i = f_i [(x_i-x_c)^2 + (y_i-y_c)^2] ,
\end{equation}

\noindent with:

\begin{equation}
\sum_i f_i < 0.2 f_{\rm tot} ,
\end{equation}

\noindent in which $f_i$ is the flux at ($x_i$, $y_i$), and ($x_c$, $y_c$) is the galaxy centroid. \M20 is a measure of the {\itshape variance} of the brightest 20\% of the light, and is {\itshape anti}correlated with concentration; a single-nucleus system will have a more negative \M20 value than a double-nucleus system, for example. The asymmetry, $A$, is a measure of the mirror, or central rotational, symmetry of all the light from a galaxy \citep{abra94,cons00}. It is determined by subtracting from the original image $I_{0}$ a $180\degr$ rotated image $I_{\phi}$:

\begin{equation}\label{eq:asym}
A = \frac{\Sigma |I_{0} - I_{\phi}|}{2\Sigma|I_{0}|} .
\end{equation}

\noindent The S{\'e}rsic index \citep{sers63} is defined in the radial light intensity profile: 

\begin{equation}\label{eq:sers}
I(r) \propto e^{-\kappa\left(\frac{r}{r_e}\right)^{1/n}} ,
\end{equation}

\noindent and is also called the concentration, or curvature, index. Finally, the effective radius, $r_{e}$, is the radius that encloses half of the total light emitted by the object. 

We adopt the $G$, \M20, and $A$ statistics as they have been used in many previous studies, and because their values have been compared against results from numerical simulations to calibrate morphological classifications \citep{lotz08a}. We include the S{\'e}rsic index and effective radius as they contain complementary information to the Gini coefficient. Both $n$ and $G$ are measures of central concentration, but $n$ also depends on profile shape, in that increasing $n$ gives a brighter center and a shallower falloff at large radii. Effective radius gives an estimate of the absolute size of an object. We use a S{\'e}rsic profile to fit our sample rather than e.g. a Nuker profile \citep{hern90,lauer95} since each object covers a small number of pixels; fitting a geometric mean is thus more meaningful than considering minor and major axes separately. 

To perform the $G$, \M20, and $A$ measurements we follow the approaches taken in \citet{lotz04} and \citet{petty09}. We first subtract an average sky flux from each frame, computed using the IRAF task imexamine. We then calculate the total flux within 1.5 times the Petrosian radius ($r_p$). Photometric uncertainties were estimated by changing the $H$-band center by up to 5\,\arcsec\ in random directions about the best-fit galaxy centroid, measuring $G$, \M20, and $A$ using the same radius, and calculating the standard error from 1000 iterations of this step. To estimate S{\'e}rsic indices and effective radii we used GALFIT \citep{peng02} to fit two-dimensional profiles of the form in Equation \ref{eq:sers}, following the approaches of previous authors \citep{ravi06,petty09}.

The light profiles of two objects -- W0514 and W2337 -- suggested that PSF subtraction might unveil more details of the host galaxy, but our attempts to do so were unsuccessful. When we included a PSF, generated using the Tiny Tim tool, as an additional component within GALFIT when fitting these two objects we could not extract stable host galaxy parameters, and the fit quality did not improve. We therefore conclude that an emerging AGN in the optical is unlikely in any of our sample, but we cannot exclude that AGN light affects the central few pixels.

\begin{figure*} 
\begin{center}
 \includegraphics[width=5.4cm,angle=0]{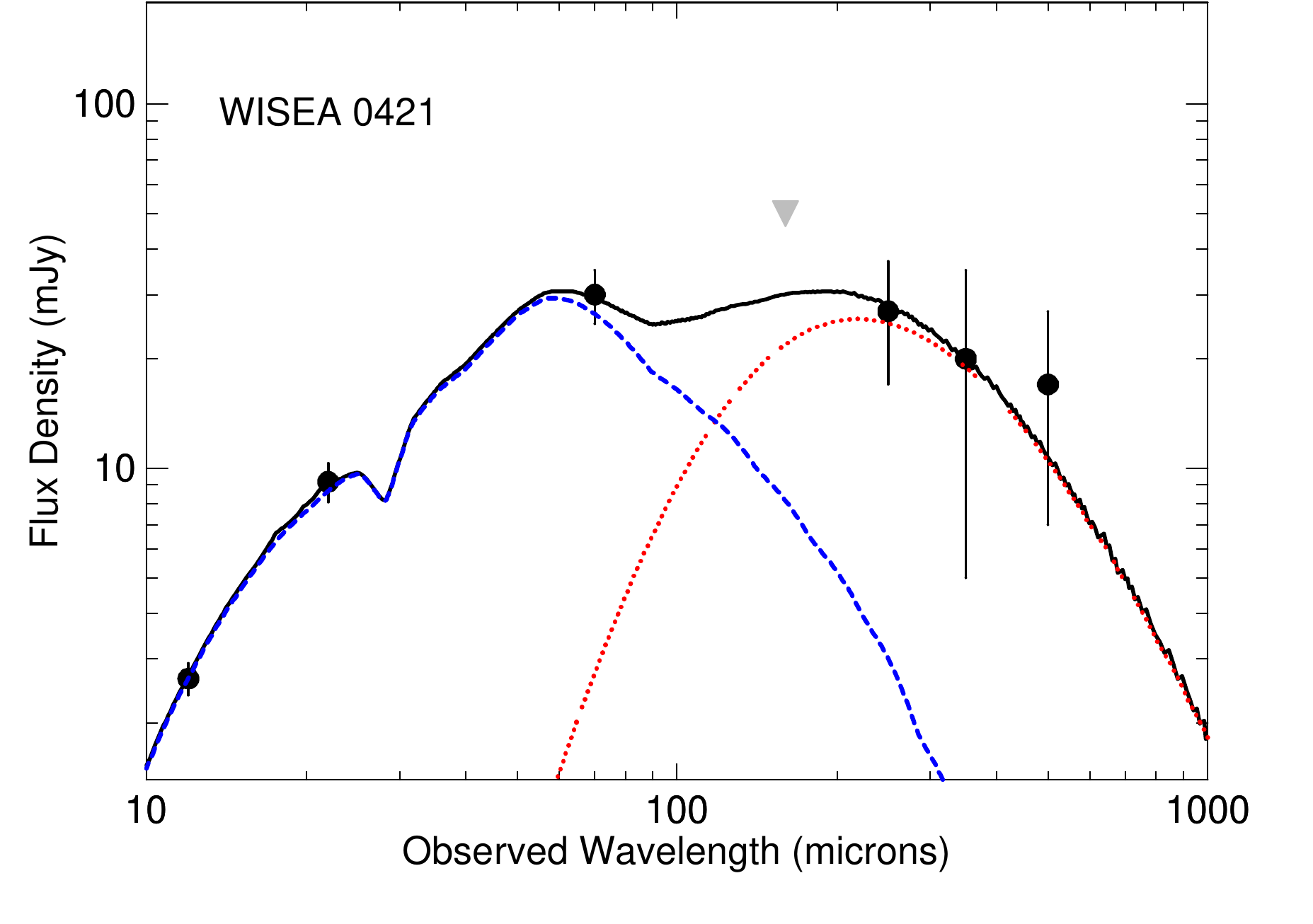} 
 \includegraphics[width=5.4cm,angle=0]{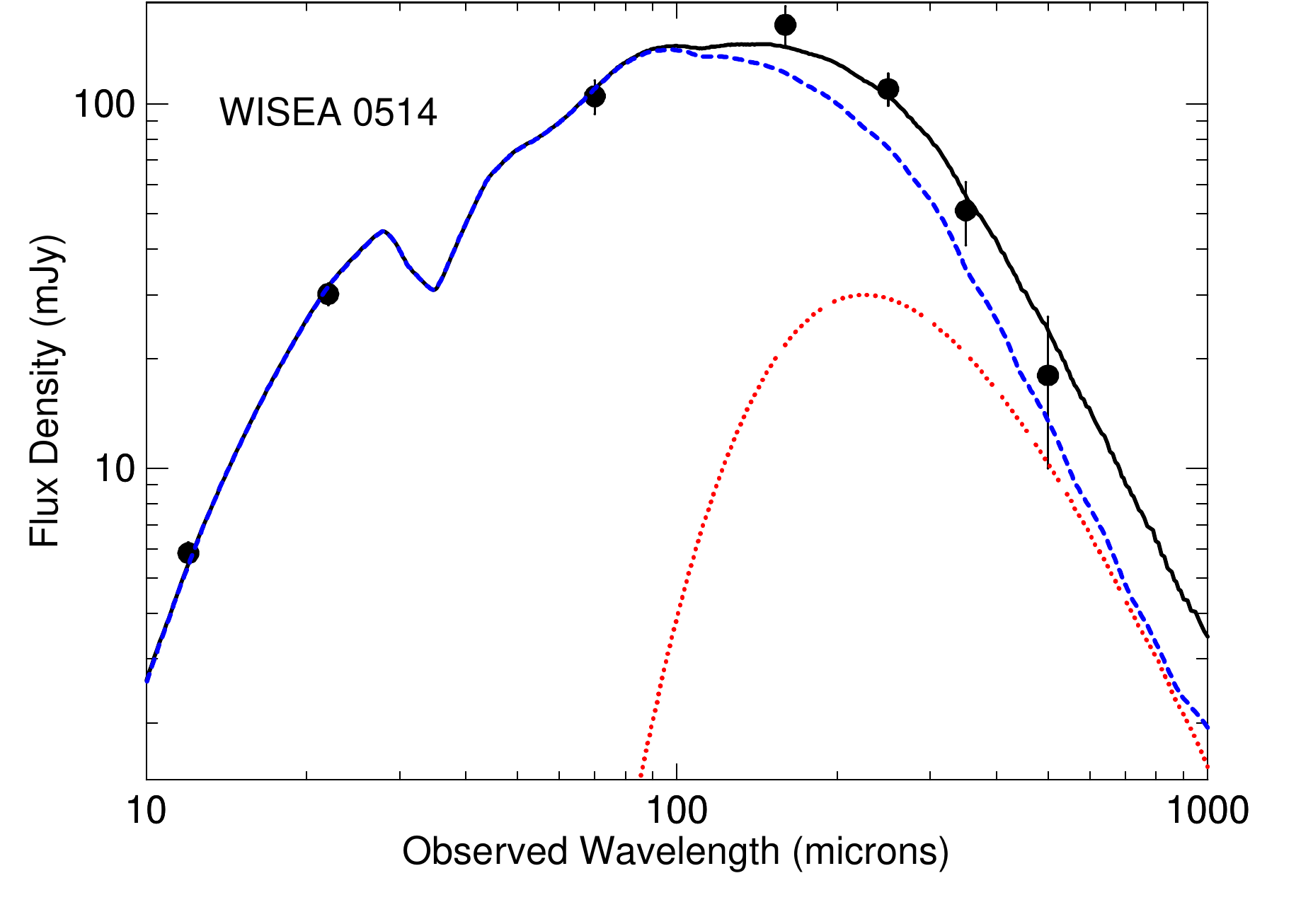} 
 \includegraphics[width=5.4cm,angle=0]{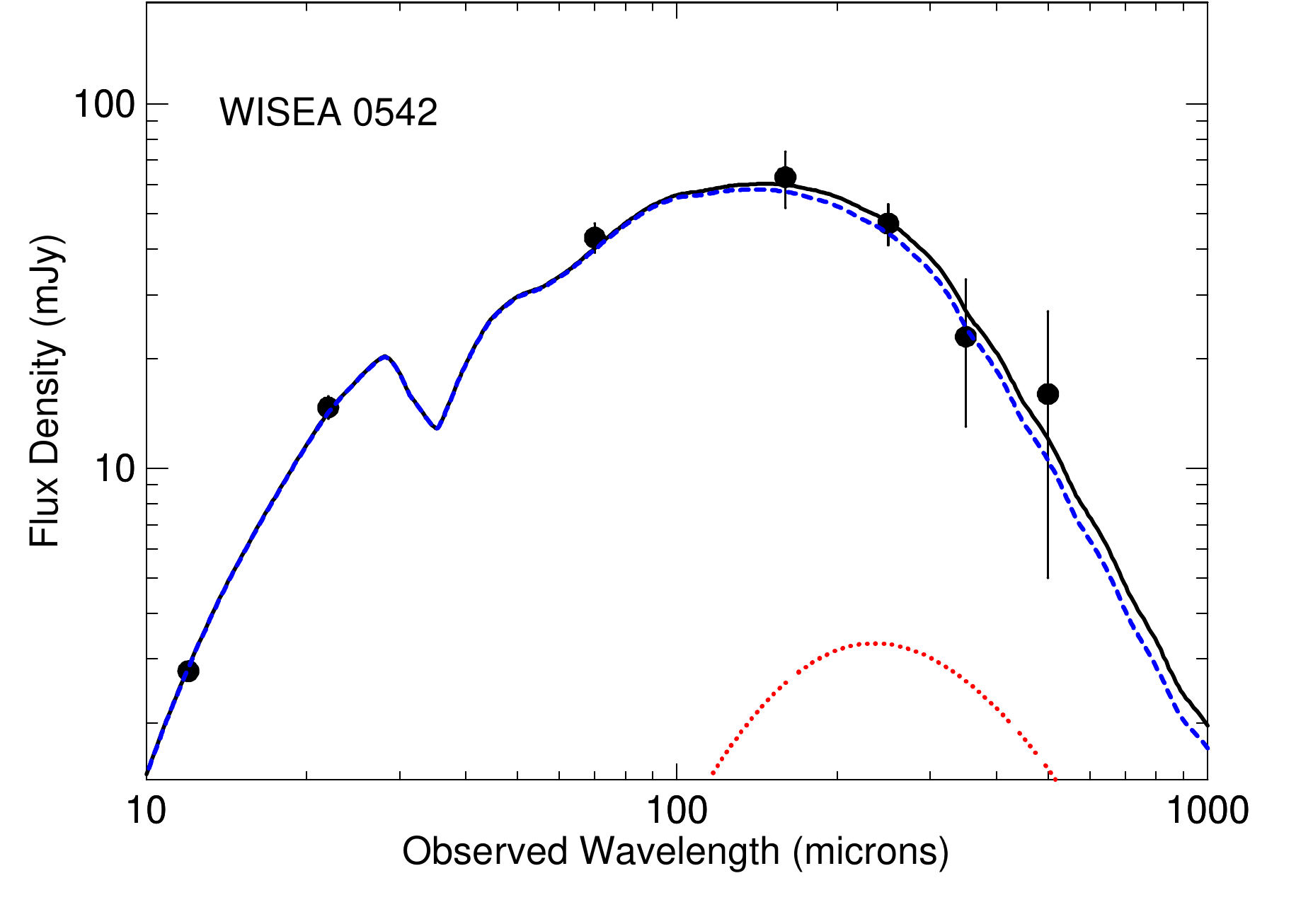} 
 \includegraphics[width=5.4cm,angle=0]{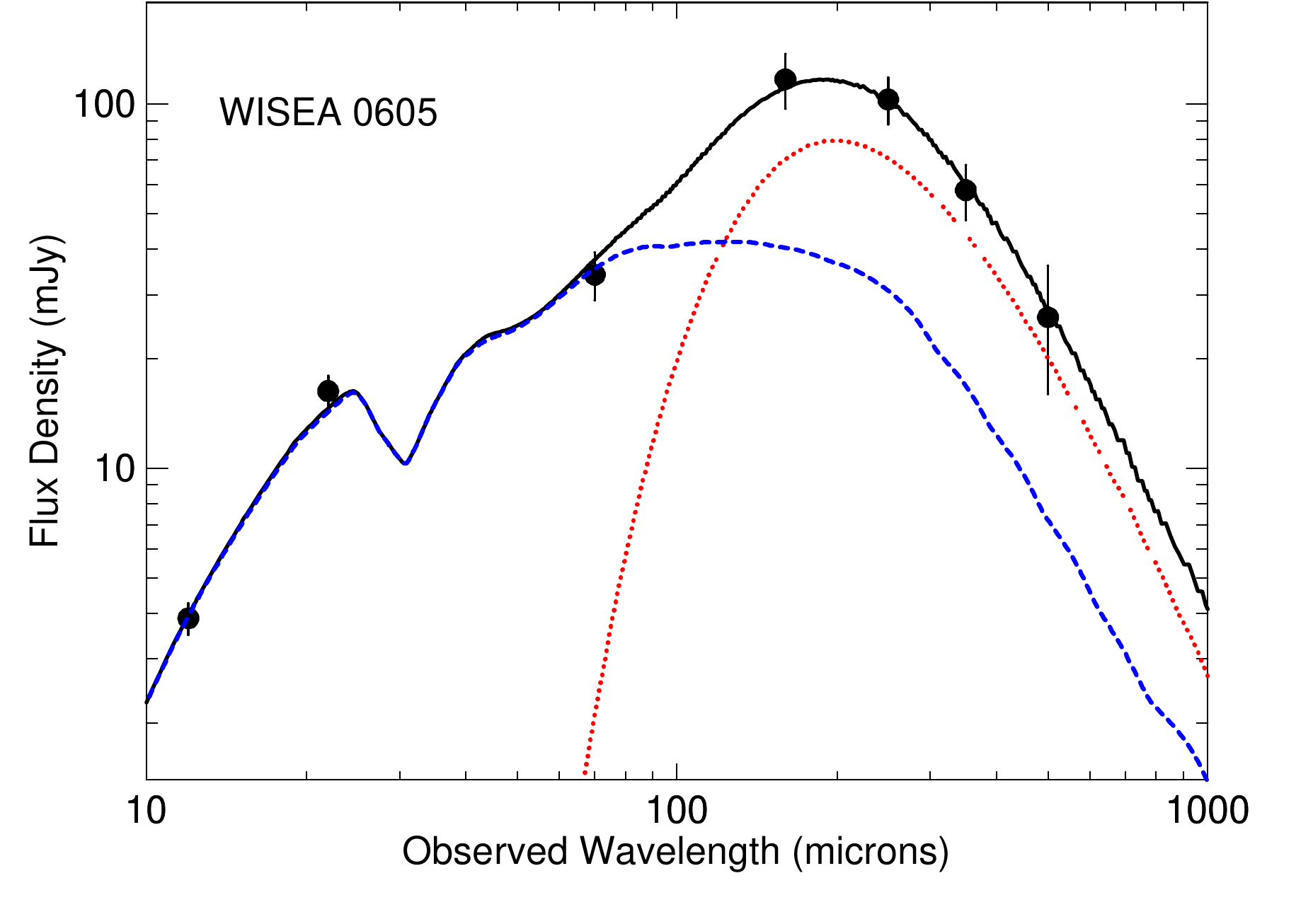} 
 \includegraphics[width=5.4cm,angle=0]{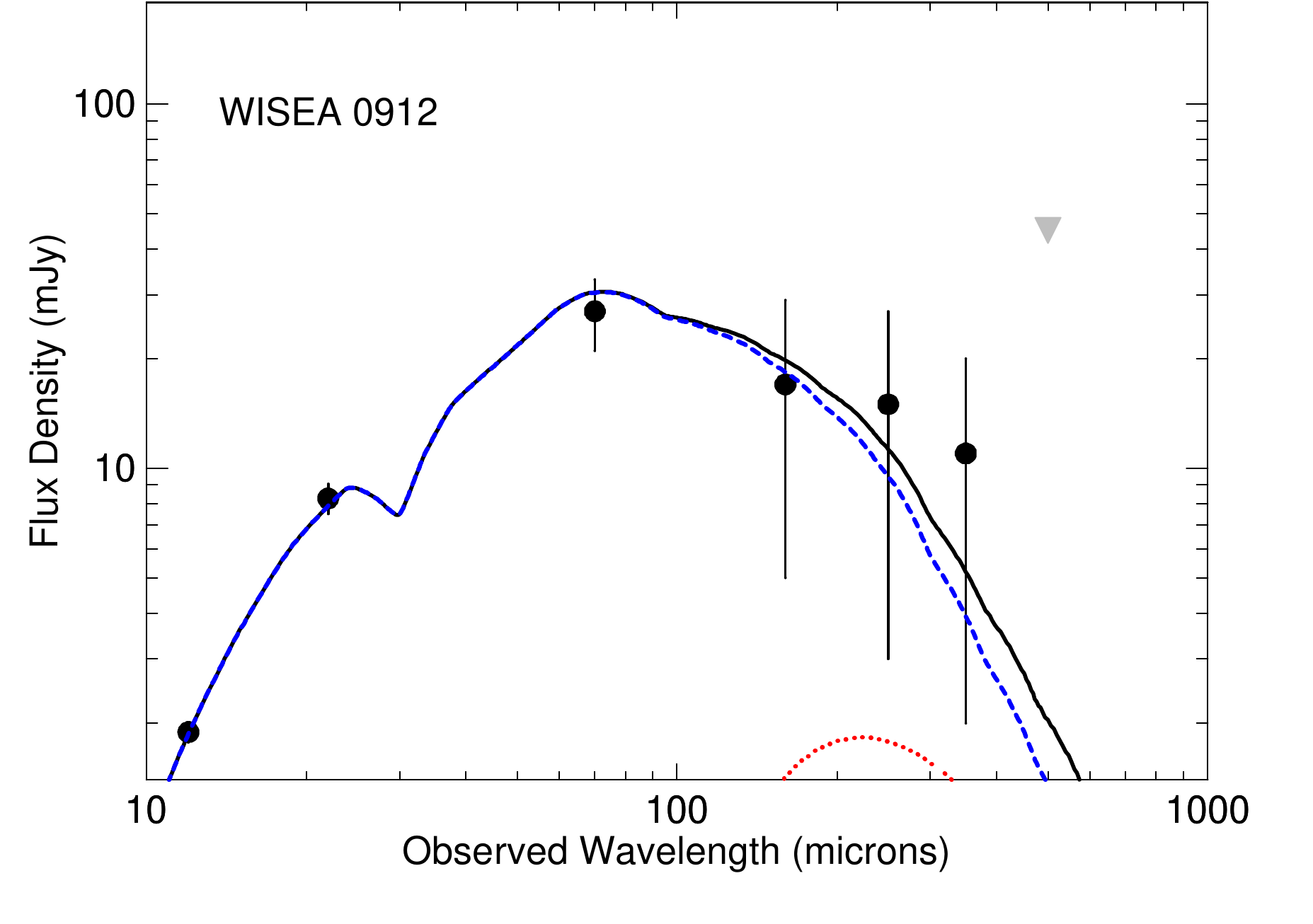} 
 \includegraphics[width=5.4cm,angle=0]{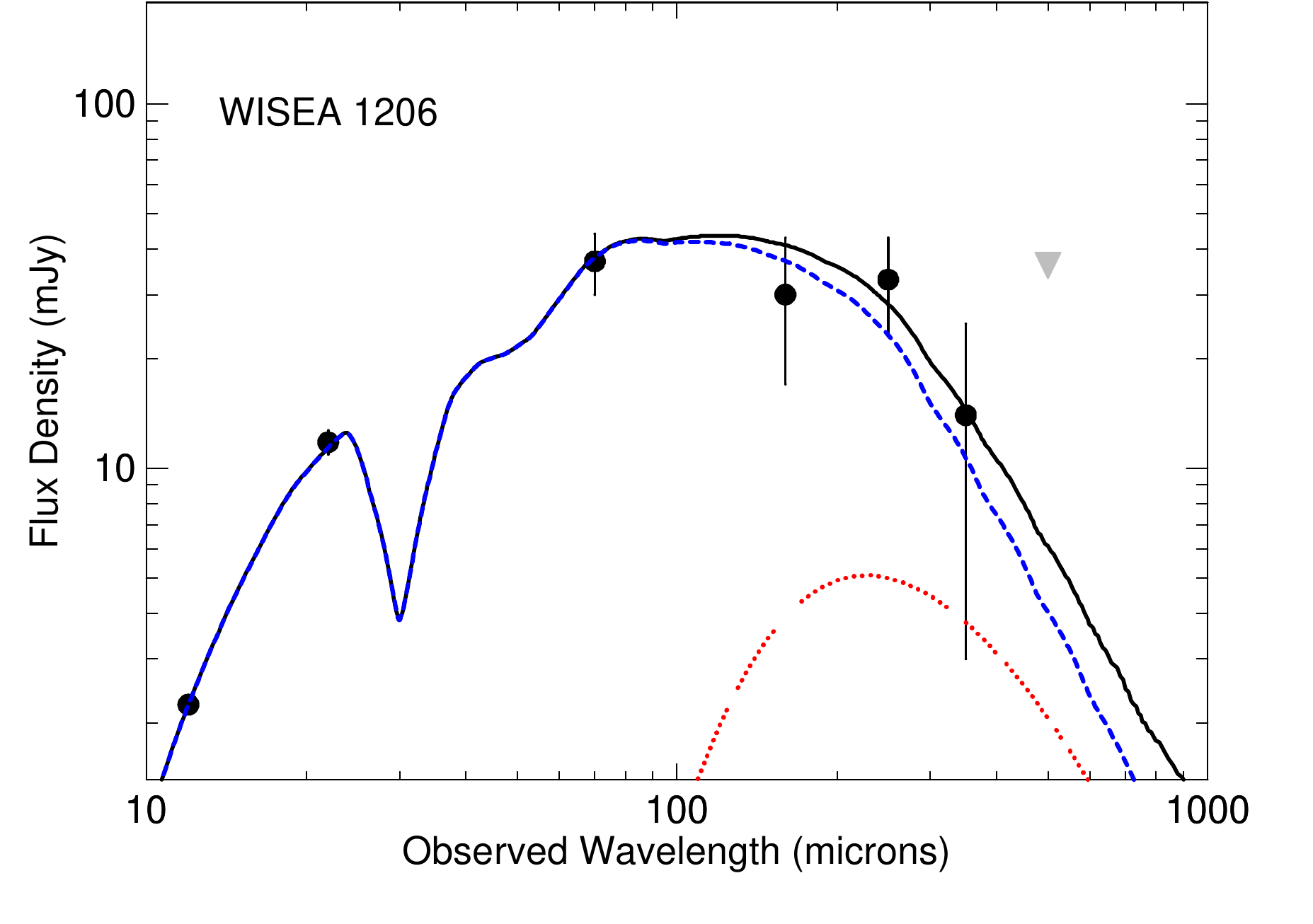} 
 \includegraphics[width=5.4cm,angle=0]{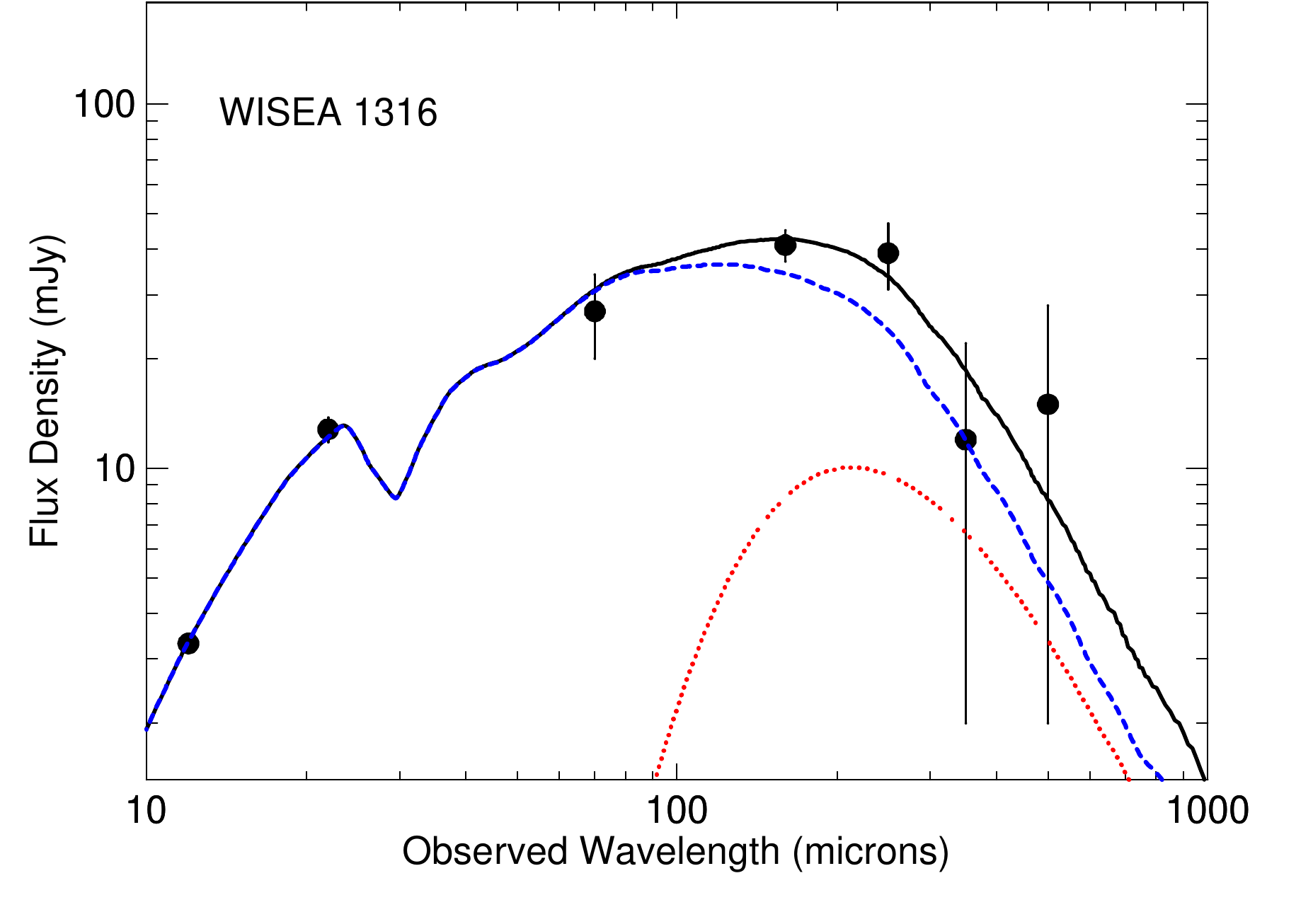} 
 \includegraphics[width=5.4cm,angle=0]{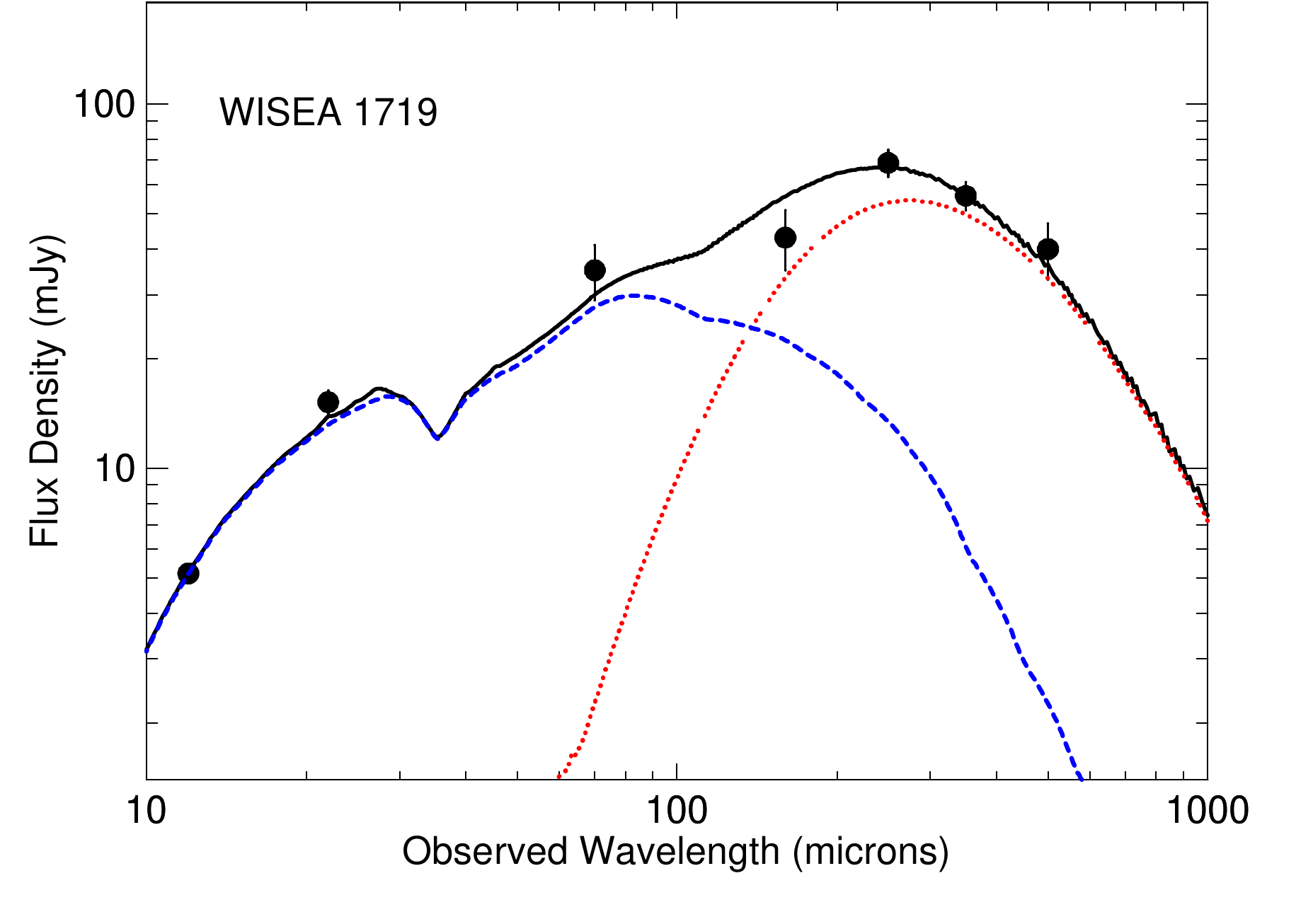} 
 \includegraphics[width=5.4cm,angle=0]{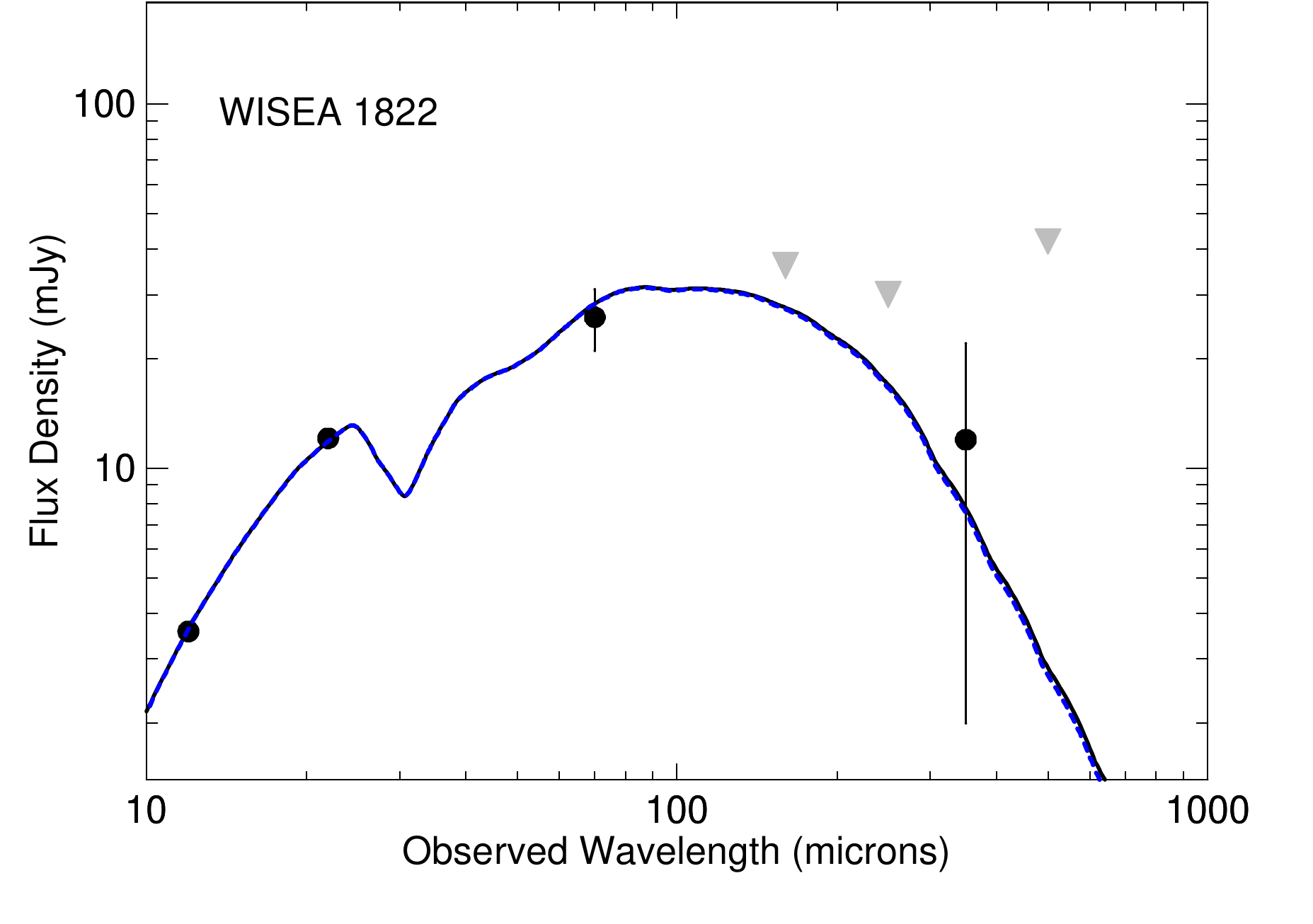} 
 \includegraphics[width=5.4cm,angle=0]{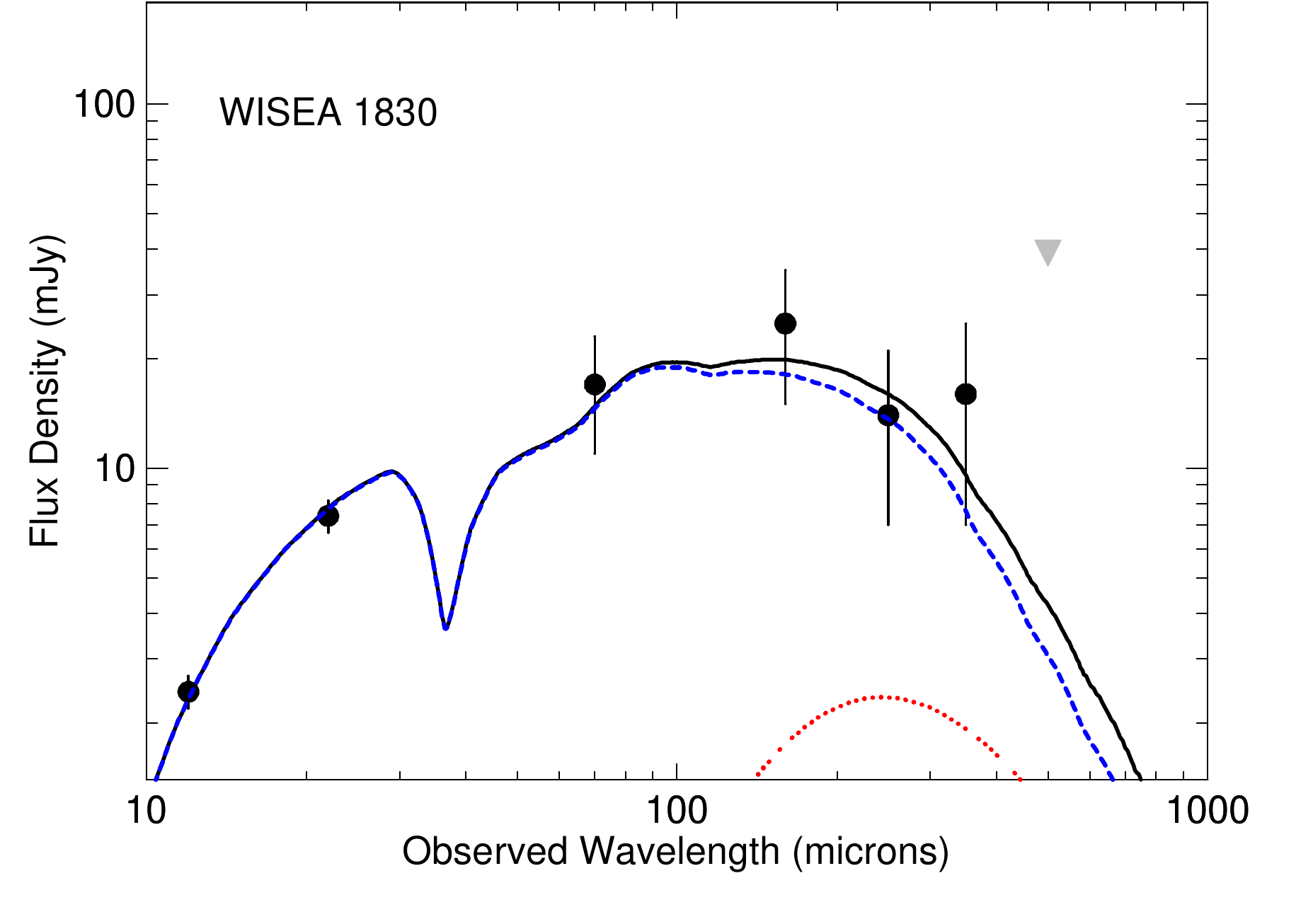} 
 \includegraphics[width=5.4cm,angle=0]{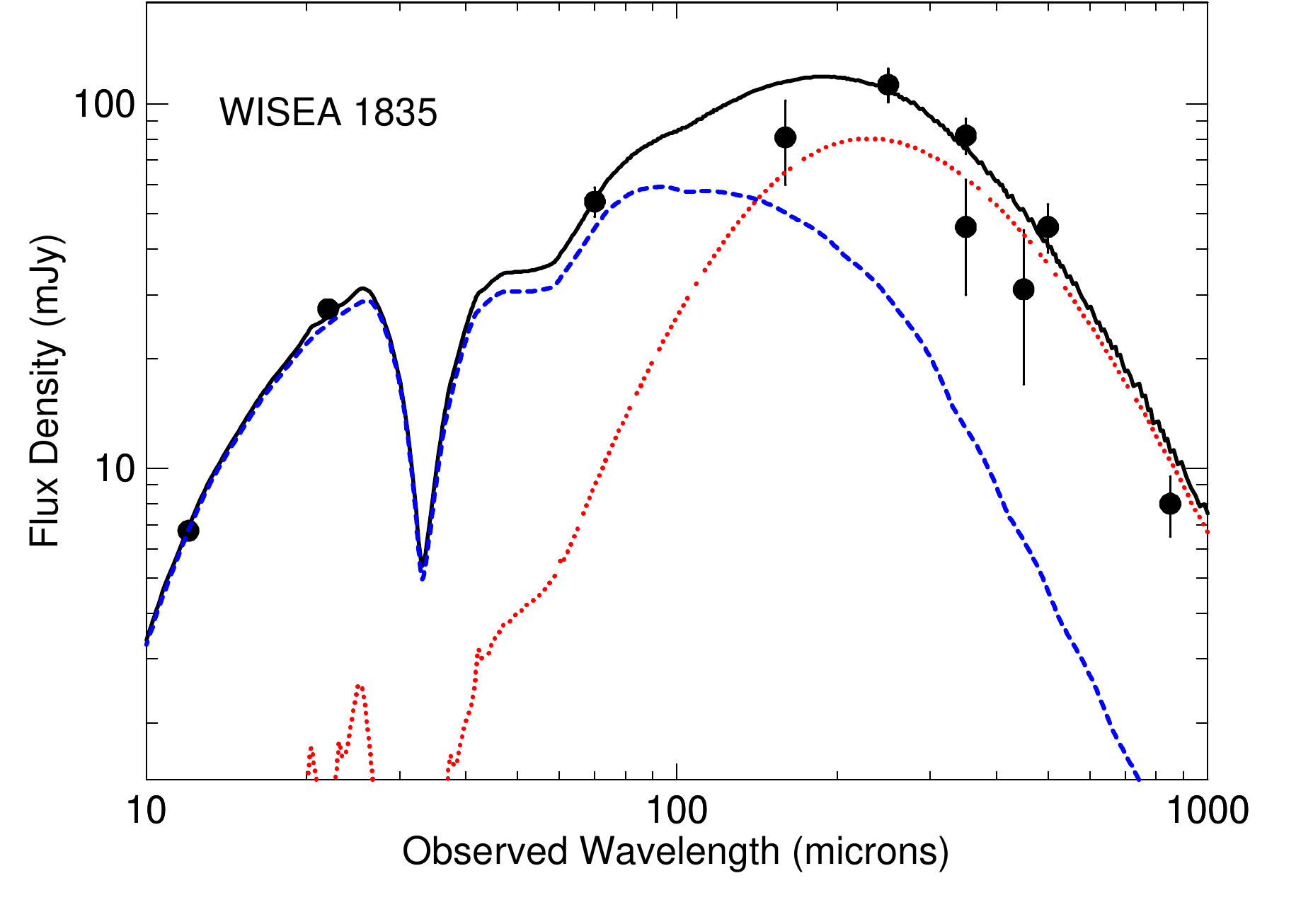} 
 \includegraphics[width=5.4cm,angle=0]{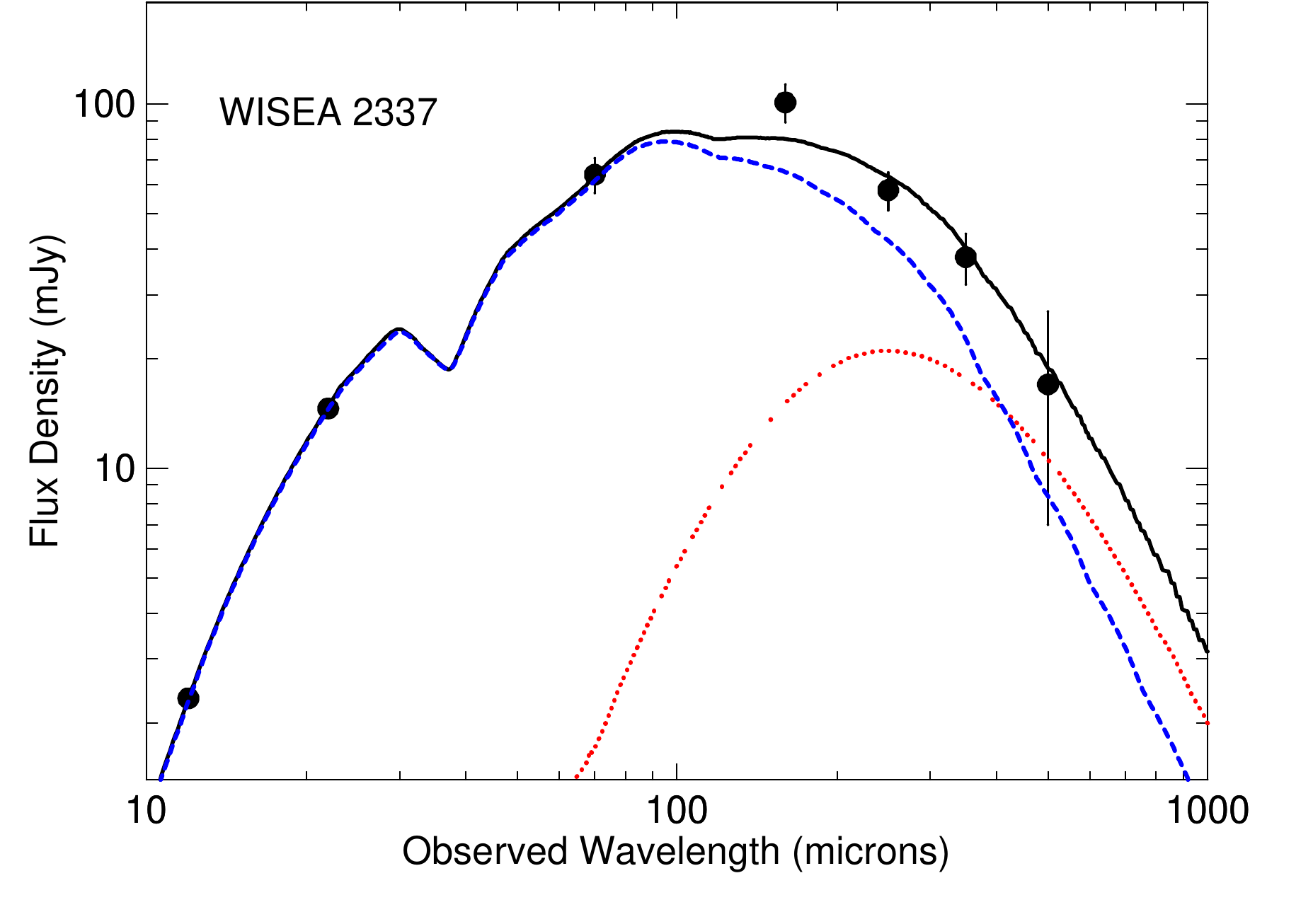} 
\caption{The mid- to far-infrared observed-frame SEDs of our sample. In each panel the black line is the combined best-fit model, the blue dashed line is the AGN component, and the red dotted line is the starburst component. Photometry is presented as $3\sigma$ upper limits (grey triangles) if detections are below $1\sigma$ significance. The ground-based photometry from Table \ref{table:basic} is also plotted. We show only the SEDs at $>4.5\mu$m as our fitting approach does not include a host galaxy component. Nevertheless, the fit is always consistent with the shorter wavelength data, either as fluxes or upper limits.}
\label{fig:seds}
\end{center}
\end{figure*}

\subsection{Infrared Emission}
To quantify the origin - AGN activity or star formation - of the infrared emission, we fit radiative transfer models to the WISE, PACS, and, where available, sub-mm and mm-wave photometry. We do not fit to data at observed-frame $4.5$\,$\mu$m and shorter wavelengths due to the possibility of host galaxy contamination, but we do include these data as limits. 

We assume that the infrared emission arises from a single episode of AGN activity, and/or star formation. We then fit the infrared data simultaneously with two grids of pre-computed radiative transfer models; one for AGN \citep{efst95,efst13} and one for starbursts \citep{efst00}. A model set for old stellar populations is not included since it is likely that the infrared emission redward of 4.5\,$\mu$m comes predominantly from obscured, luminous activity, with a negligible contribution from unobscured, older stars. These models have been used previously in \citet{ver02,farrah02,farrah03,farrah12,efst13}. The AGN models assume the dust geometry is a smooth tapered disk whose height, $h$, increases linearly with distance, $r$, from the AGN until it reaches a constant value. The dust distribution includes multiple species of varying sizes, and assumes the density distribution scales as $r^{-1}$. The starburst models combine the population synthesis code of \citet{bru03} with a prescription for radiative transfer through dust that includes the effects of small dust grains and polycyclic aromatic hydrocarbons (PAHs, \citealt{efst09}). In total there are 1680 starburst models and 4212 AGN models.

Both model sets have several free parameters (e.g. inner half-opening angle for the AGN, age and initial optical depths of the molecular clouds for the starbursts), which we lack the data to constrain. Instead, we use the full model sets to obtain a realistic estimate of the uncertainties on the luminosities by using all possible combinations of SED fits for each object to construct a weighted probability distribution for the total, AGN, and starburst luminosities. The only constraint we impose is on the AGN model set; that the viewing angle is greater than the torus half opening angle, as measured from pole-on, so that the broad line region is not visible in direct light. This reduces the number of AGN models to 2064. This approach assumes that all models exist in the high redshift infrared-luminous galaxy population, and that they are comparably likely. This assumption lacks strong supporting evidence, but the model sets do span the properties found in lower redshift populations. Moreover, our approach is superior to simply normalizing a single template, or fitting a small number of templates, as such an approach is effectively a small set of delta functions in the same parameter space.

\section{Results \& Analysis}\label{results}
The HST images are presented in Figure \ref{fig:imagesa}. We do not present the WISE and {\itshape Herschel} images as they are in all cases consistent with point sources, see \S\ref{ssircols}. The $H$-band, {\itshape Spitzer}, WISE, PACS and SPIRE fluxes are presented in Table \ref{table:basic}. The morphological parameters and infrared luminosities are presented in Table \ref{table:lummorph}. The infrared SEDs are presented in Figure \ref{fig:seds}.

\begin{figure*} 
\begin{center}
\includegraphics[width=20cm,angle=0]{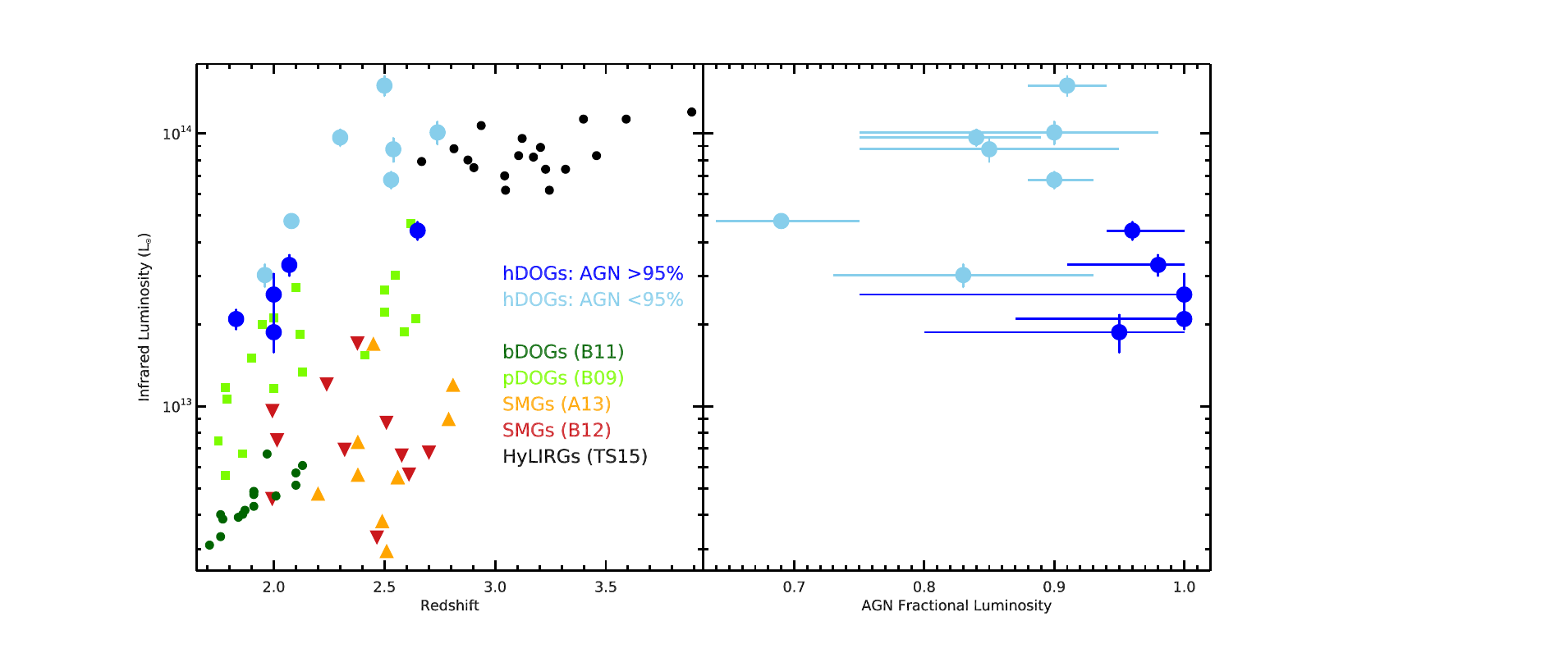}  
\caption{The infrared luminosities of the $h$DOGs as functions of (left) redshift and (right) AGN fractional luminosity. The left panel also includes comparison populations from the literature (\S\ref{compops} \& \citealt{tsai15}).}
\label{fig:lirvszfrac}
\end{center}
\end{figure*} 

\subsection{Comparison Samples}\label{compops}
In the following, we make comparisons to five samples from the literature. First are the `bump' DOGs ($b$DOGs) and `power law' DOGs ($p$DOGS, \citealt{buss09,buss11,mel12}). The $b$DOGs have mid-infrared continua showing an opacity minimum arising from H$^{-}$ in stellar atmospheres, while the $p$DOGs have mid-infrared continua showing an AGN power law. The $b$DOGs have fainter $22\,\mu$m flux densities and less extreme $R - [22]$ colors than the $p$DOGs. Second are two samples of sub-mm selected galaxies (SMGs), one from \citet{buss11} and one from \citet{agui13}. The SMGs from  \citet{buss11} are the 25 SMGs with redshifts in the range $0.7 <z< 3.4$, originally presented in \citet{swin10} and selected from the spectroscopic catalogue of \citet{chapman05}. The parent sample are the radio-identified subset of sources detected at $850\,\mu$m in surveys with SCUBA. The SMGs from \citet{agui13} lie approximately in the same redshift range as our sample, and include both unlensed and lensed sources, and sources selected at both $850\,\mu$m and $1.2$\,mm. While their selection is heterogeneous compared to that of \citet{buss11}, they publish asymmetries, so we include them for comparison. The final comparison sample is the sample of M$_{*}>10^{11}$M$_{\odot}$ galaxies at $1.7<z<3.0$ from the GOODS NICMOS Survey (GNS, \citealt{bui08,cons11}). This sample comprises massive galaxies with infrared luminosities much lower than any of the other comparison samples, and is not subdivided into quiescent and star forming subsamples.

Each comparison sample includes only a subset of the morphological parameters we consider. In summary, the samples used for morphological comparisons are:

\begin{itemize}
\item $h$DOGS  (\#1, our sample):       $G$, \M20, $A$, $r_{e}$, and $n$
\item $b$DOGs  (\#2, \citealt{buss11}): $G$, \M20, $r_{e}$, and $n$
\item $p$DOGs  (\#3, \citealt{buss09}): $G$, \M20, $r_{e}$, and $n$
\item B12 SMGs (\#4, \citealt{buss11}): $G$, \M20, $r_{e}$, and $n$
\item A13 SMGs (\#5, \citealt{agui13}): $G$, \M20, $A$ 
\item GNS      (\#6, \citealt{bui08}):  $G$, \M20, $A$, $r_{e}$, and $n$
\end{itemize}

\noindent All of the comparison samples were observed by HST at rest-frame wavelengths closely matched to our sample (we do not consider the minor difference in resolution between NICMOS-NIC2 and redrizzled WFC3). In all cases we include only those objects within approximately the same redshift range as our sample. At $z=2$ the spatial scale is $8.37$\,kpc\,arcsec$^{-1}$. For these reasons we do not include comparisons to objects within the COSMOS field observed with ACS, or to low redshift ULIRGs and Hubble sequence galaxies \citep{lotz04}, as the finer spatial resolution and different rest-frame wavelengths could lead to invalid comparisons.

 \begin{deluxetable}{lccccc}\label{table:probsim}
 \tablewidth{0pt}
 \tabletypesize{\scriptsize}
 \tablecaption{Probabilities of similarity between our sample and the comparison samples}
 \tablehead{
 \colhead{Sample} & \colhead{$G$}  & \colhead{\M20} & \colhead{$A$} & \colhead{$r_{e}$} & \colhead{$n$} \\  
 }                                                                     
 \startdata                                                                      
\#2: $b$DOGs      & 0.57        & 0.23      & --      & 96            &  35     \\
\#3: $p$DOGs      & $<0.1$      & 9.7       & --      & 79            &  19     \\
\#4: B12 SMGs     & 1.1         & 2.4       & --      & 99            &  99     \\
\#5: A13 SMGs     & 66          & 55        & 35      & --            &  --     \\
\#6: GNS galaxies & 0.12        & 2.2       & 99      & 82            &  99     \\
 \enddata                                      
 \tablecomments{The percentage probabilities that the $h$DOGs have a similar distribution to literature samples in each morphological parameter. The probabilities are computed using the approach described in \S\ref{compops}. In the text, these probabilities are denoted $P_{1,x}^{y}$, where $x$ is the population being compared against the $h$DOGs, and $y$ is the morphological parameter in question.}
 \end{deluxetable}

To make quantitative comparisons between populations, we employ two-sample Bayesian hypothesis testing using a nonparametric Polya tree prior \citep{holmes15}. This approach is superior to the traditional Kolmogorov-Smirnov test (or other frequentist approaches) as it gives the probability {\itshape for} the null hypothesis that the two populations are identical, rather than the probability of obtaining the same or a more extreme result {\itshape assuming} that the two populations are identical. The probabilities are given in Table \ref{table:probsim}.

\subsection{Infrared luminosities and colors}\label{ssircols}
In the WISE images our sample are all consistent with point sources at $12$ and $22\,\mu$m. The PACS and SPIRE images are also consistent with point sources. In all cases the WISE and (where detected) {\itshape Herschel} sources are centered closely on the primary HST source. This, combined with the absence of foreground objects in the spectra, and the $h$DOG selection which biases against bright foreground objects, means that significant gravitational lensing amplification of our sample is unlikely. 

The fitted total rest-frame infrared luminosities of our sample span $1.8\times10^{13}$\,L$_{\odot}$ to $1.4\times10^{14}$\,L$_{\odot}$, making them among the most luminous objects in the Universe (see also \citealt{Bridge2013,tsai15}). They are much more infrared-luminous than $z<0.2$ ULIRGs \citep[e.g.][]{farrah03}, and have comparable luminosities to HLIRGs found at other redshifts that were selected at observed-frame $\lesssim100\,\mu$m \citep{row00}. They match or exceed the luminosities found for the most luminous SMGs \citep{chapman05}. Four of our sample (W0542, W1316, W1830, W1835) have infrared luminosities in \citet{fan16}; in three cases our luminosities are consistent with theirs, while for the fourth (W1835) our luminosity is a factor of two higher. This difference likely arises from our more comprehensive infrared data, and the different approaches to modelling the AGN and the starburst emission. W1835 also has a published infrared luminosity in two previous papers; our luminosity is 60\% higher than that derived by \citet{wu12}, but consistent with the (template) luminosity of \citet{jones14}.

The combination of WISE, PACS, SPIRE, and (in some cases) ground-based sub-mm and mm-wave photometry means we can constrain the fraction of the total infrared luminosity that arises from AGN ($f\mathrm{_{AGN}}$). In all cases the infrared SED fits mark our sample as luminous, obscured, AGN-dominated systems. In some cases it is possible to explain {\itshape all} the infrared emission as arising from AGN activity, though the best-fit solution usually includes some star formation, with star formation rates of up to a few hundred Solar masses per year. The AGN fractions of our sample are consistent with being higher than in the $b$DOGs, and higher or comparable to that in the $p$DOGs \citep{pope08,ass15}. 

We note three caveats to the results from the infrared SEDs. First, since the AGN models are axisymmetric, the total infrared luminosity cannot be precisely inferred simply by integrating the line-of-sight infrared luminosity over $4\pi$ steradians. However, in all cases the anisotropy correction to the AGN luminosity is a factor of two or less, and in most cases the $1\sigma$ uncertainties on the anisotropy correction encompass unity, so we do not apply them here. Second is the possibility that the bias towards hotter dust in our sample compared to other classes of DOGs is consistent with {\itshape younger} starbursts, rather than a particular AGN phase. This possibility arises since younger starbursts have a more intense interstellar radiation field, and therefore have elevated dust temperatures compared to older starbursts \citep{efst09}. We cannot, however, explore this possibility, since the lack of rest-frame mid-infrared spectra means that the constraints on the starburst ages from the model fits are weak - the 90\% confidence intervals on the starburst ages are $10-60$\,Myr or wider in all cases. Tighter constraints than this would require higher quality infrared data \citep{farrah16}. Third, we cannot exclude the possibility of a contribution to the total infrared luminosity from `cirrus' dust heated by quiescent starlight. 

Since the uncertainties on the fractional AGN luminosities are significant, we define two classes of AGN fraction for subsequent analysis. These classes are $f\mathrm{_{AGN}} < 0.95$ and $f\mathrm{_{AGN}} > 0.95$. The choice of boundary is to some extent arbitrary, and is intended to help frame the discussion, we do not adopt it out of any physical motivation. We define the classes such that one is an ``AGN composite'' class, and the other is a ``pure AGN'' (or close to) class.

In Figure \ref{fig:lirvszfrac} we plot the luminosities of our sample against both redshift and AGN fractional luminosity. There is, perhaps, a trend with redshift, with the more luminous objects lying at higher redshifts, though this could be a selection effect. There is also a hint of a trend with AGN fraction, with 5/6 objects with the lowest infrared luminosities having the highest AGN fractional luminosities. Compared to samples from the literature; our sample are more luminous than either the $b$DOGS or $p$DOGs, by factors of approximately three and ten, respectively. This is consistent with the idea that the $h$DOG selection preferentially finds obscured, extremely infrared-luminous AGN. Our sample are also more luminous than any of the SMGs in the same redshift range. Finally, we compare to the HyLIRG sample of \citealt{tsai15} (TS15). The TS15 sample was selected in the same way as ours, but comprise the most luminous 20 objects in the parent spectroscopic catalog of $h$DOGs. Compared to the TS15 sample, our sample are at lower redshifts than the TS15 sample, and slightly lower, on average, infrared luminosities, though at $z\gtrsim2.2$ their luminosities are comparable. 

\begin{figure*} 
\begin{center}
\includegraphics[width=20cm,angle=0]{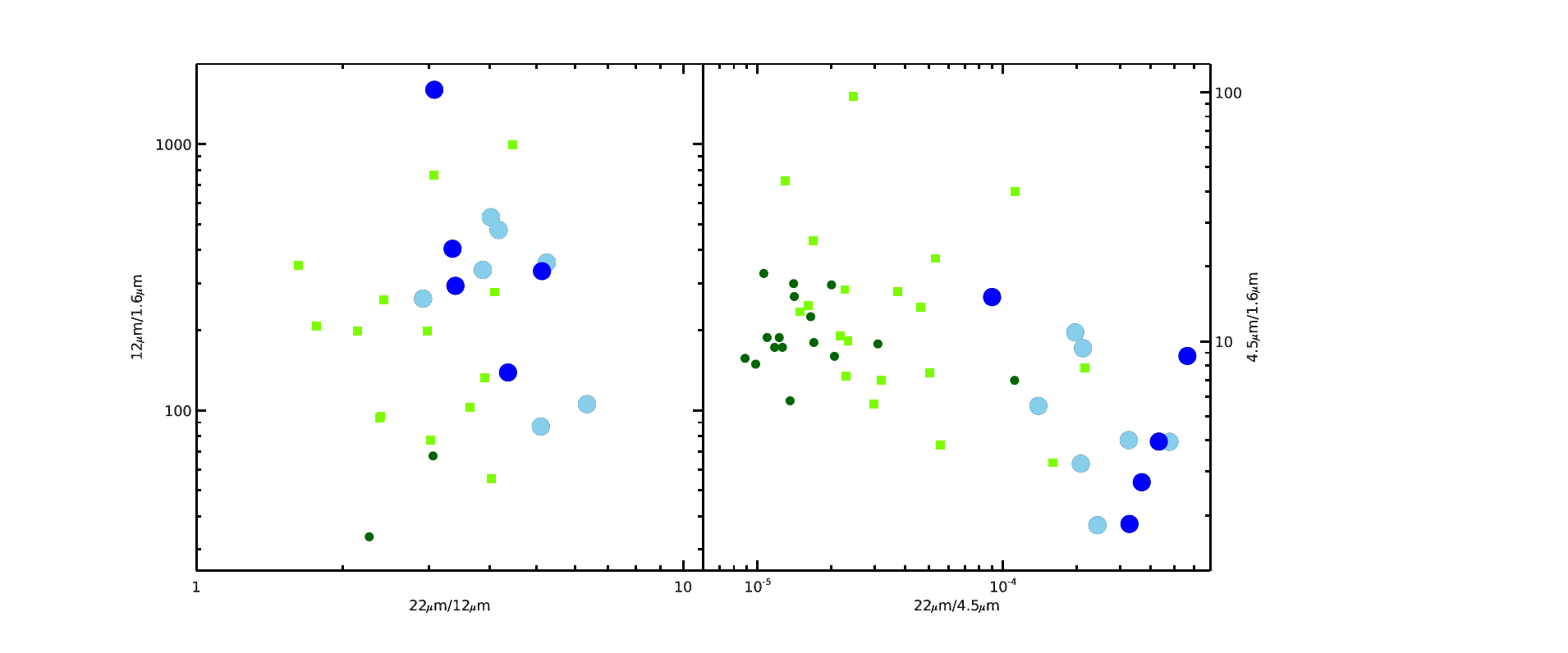} 
\caption{Two color-color plots for our sample, with the $p$DOGs and the $b$DOGs (\S\ref{compops}) also shown. Only those objects with detections in all relevant bands are plotted. See Figure \ref{fig:lirvszfrac} for a key to the points.}
\label{fig:ircolors}
\end{center}
\end{figure*}

Turning to the infrared colors; in Figure \ref{fig:ircolors} we consider two color-color plots: $f_{22}/f_{12} - f_{12}/f_{1.6}$ and $f_{22}/f_{4.5} - f_{4.5}/f_{1.6}$ (the TS15 HyLIRG sample are not plotted, because 1.6 \micron \, photometry is not available for this sample). Starting with the $f_{22}/f_{12} - f_{12}/f_{1.6}$ plot; our sample have similar $f_{12}/f_{1.6}$ colors to the $p$DOGs, consistent with both populations having similar AGN to host galaxy luminosity ratios, at least at 12$\mu$m. However, our sample have significantly redder $f_{22}/f_{12}$ colors than the $p$DOGs or the $b$DOGs. This is consistent with the AGN in the $h$DOGs being more obscured than in other classes of DOG. In the $f_{22}/f_{4.5} - f_{4.5}/f_{1.6}$ plot, we see a clear separation; the $h$DOGs have higher $f_{22}/f_{4.5}$ ratios than the $p$DOGs, which are themselves higher than the $b$DOGs, consistent with a rising contribution from an obscured AGN for the same host galaxy mass. However, the $h$DOGs have lower $f_{4.5}/f_{1.6}$ ratios than either the $p$DOGs or $b$DOGs, which are similar to each other. This is consistent with a host galaxy that is one or more of younger, less massive, or less obscured, and/or that there is some contribution to the optical light from the central AGN.

\subsection{Morphologies}\label{qmethods}
The WFC3 images show a range of morphologies. One object, W0542, has a symmetric light profile, appears undisturbed and has no close companions. Three objects (W1206, W1316, W1835) show evidence for a disturbed, irregular light profile but do not clearly have close companions. The remaining eight objects all show disturbed profiles with what appear to be one or more close companions. In all cases the spatial extents of the sources lie approximately in the range 10-25\,kpc. No object shows evidence for significant gravitational lensing (see also \citealt{wu14}).

For an initial morphological classification, we use boundaries in the $G$ -- \M20 -- $A$ plane. In the $G$ -- \M20 plane we use the boundaries determined by \citet{lotz08b} for mergers:

\begin{equation}\label{eq:gm20a}
G > -0.14 M_{20} + 0.33 
\end{equation}
 
\noindent early-type:

\begin{equation}\label{eq:gm20b}
G \leq -0.14 M_{20} + 0.33 {\rm ; \ and \ } G > 0.14 M_{20} + 0.80
\end{equation}

\noindent and late-type systems:

\begin{equation}\label{eq:gm20c}
G \leq -0.14 M_{20} + 0.33 {\rm ; \ and \ }  G \leq 0.14 M_{20} + 0.80
\end{equation}
 
\noindent as determined from comparisons to galaxies at $z\sim0.3$ observed in the rest-frame B-band, at an effective resolution of 0.62\,kpc (see also \citealt{lotz04,lotz08a}). To classify mergers vs non-mergers in the $G$ -- $A$ plane, \citet{con03} and \citet{lotz04} propose a boundary of:

\begin{equation}\label{eq:mergerA}
G > -0.4 A + 0.68  {\rm ; \ or \ }  A \geq 0.35
\end{equation}

\begin{figure*} 
\begin{center}
\includegraphics[width=8cm,angle=0]{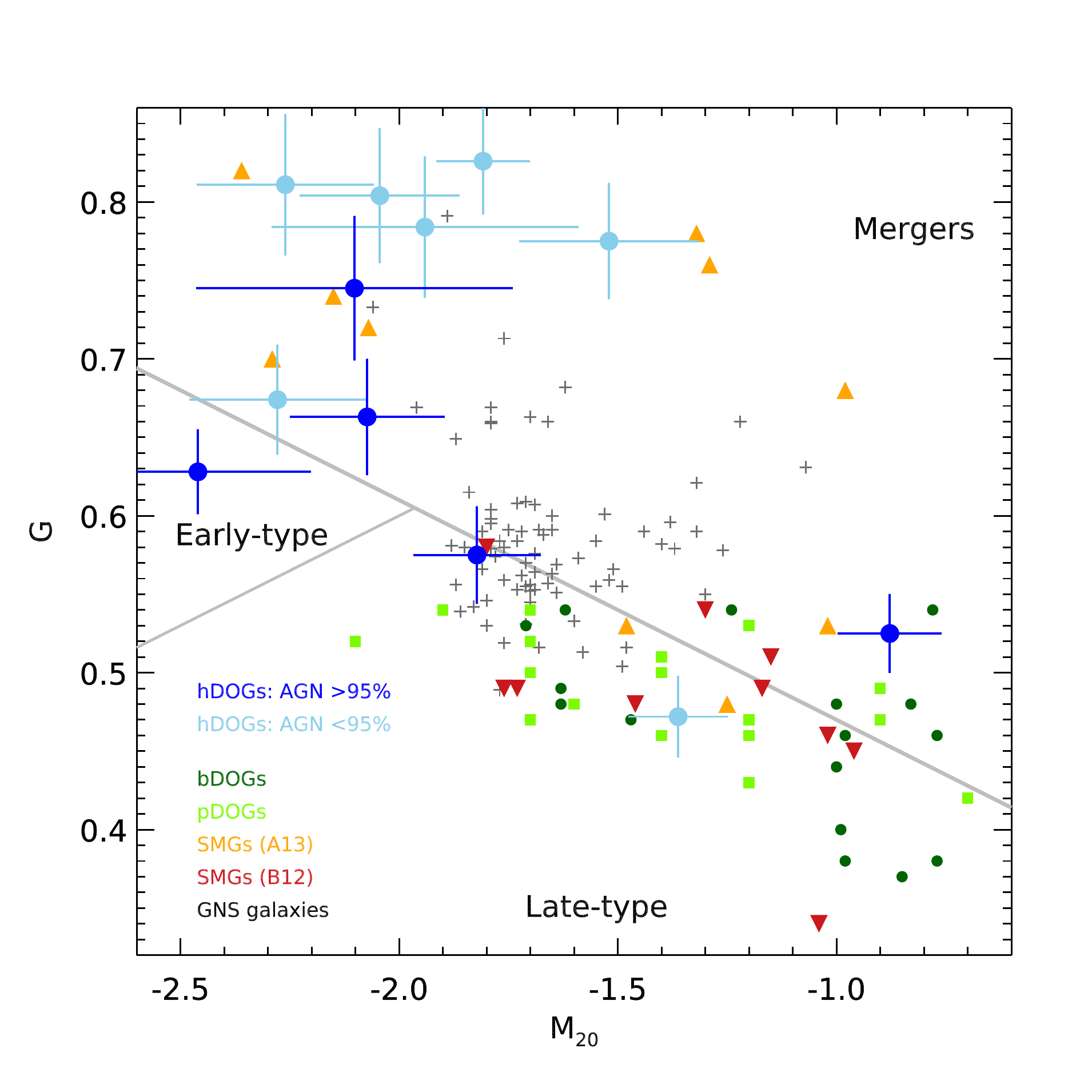} 
\includegraphics[width=8cm,angle=0]{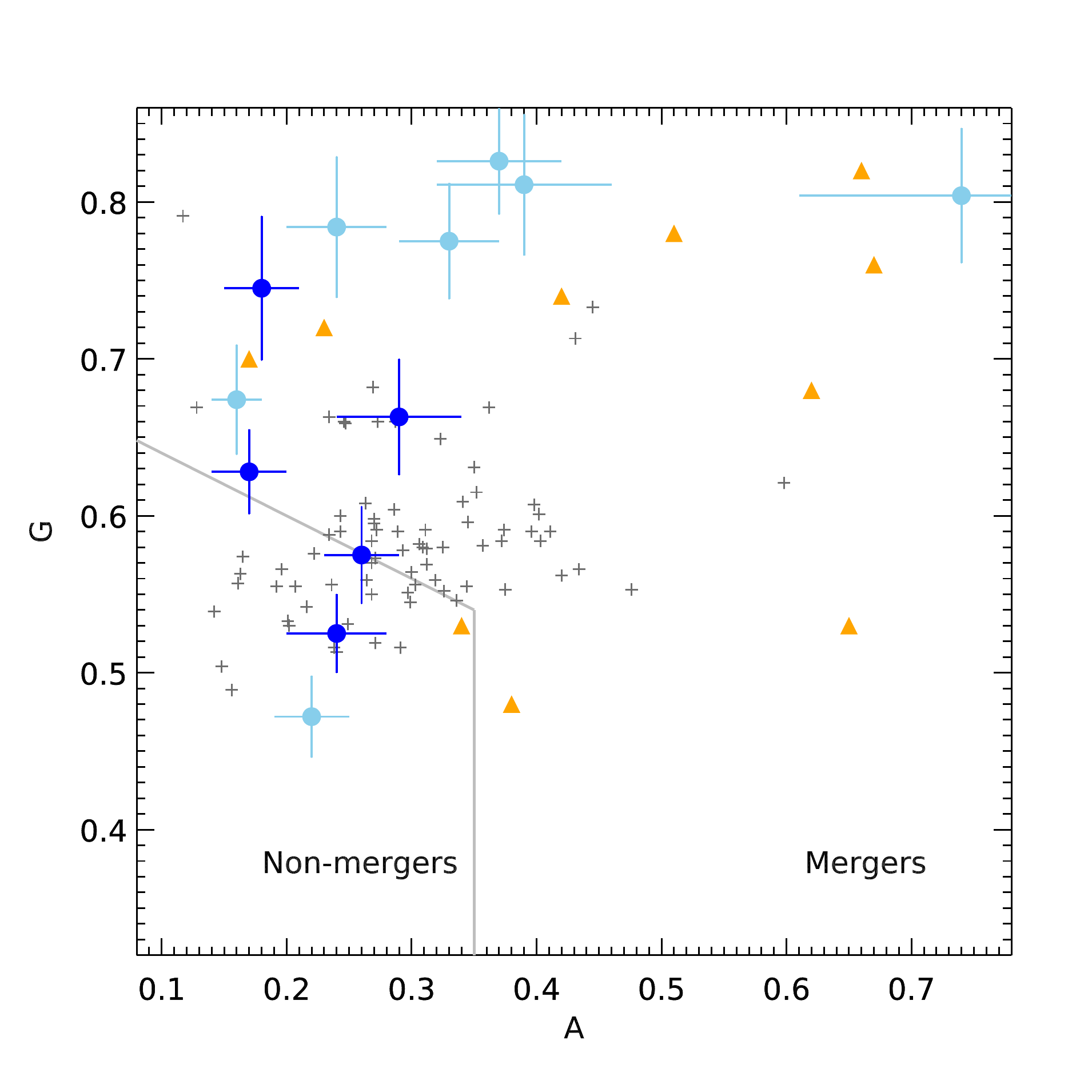} 
\includegraphics[width=8cm,angle=0]{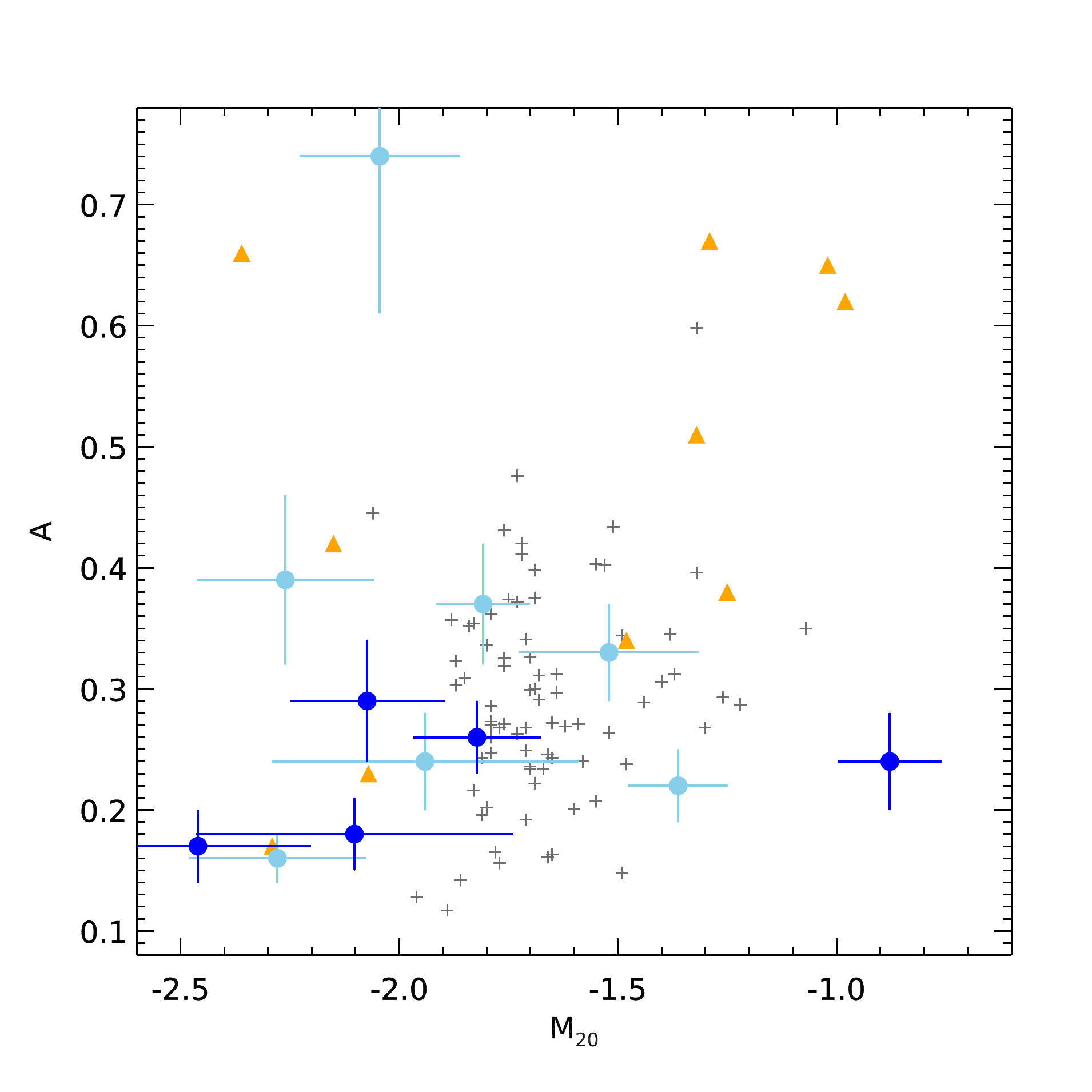} 
\caption{The locations of our sample in the $G$-\M20-$A$ planes, including comparison populations (\S\ref{compops}), and morphological boundaries (\S\ref{qmethods}). As described in \S\ref{submorph} $G$ is a measure of how concentrated the light is in an image (higher $G$ corresponding to more concentrated light), \M20 is a measure of the {\itshape variance} of the brightest 20\% of the light (more negative \M20 corresponding to less variance), and $A$ is a measure of the mirror symmetry of all the light from a galaxy (higher $A$ corresponding to greater asymmetry).}
\label{fig:gm20a}
\end{center}
\end{figure*}

\noindent Our sample, and these boundaries, are shown in Figure \ref{fig:gm20a}. 

The use of these boundaries for our sample comes with four caveats. First, the spatial resolution at which the boundaries were determined, at $\sim0.6$\,kpc, is finer than the 1.26\,kpc resolution of WFC3/IR at $z=2$. The general effect of this is to lower $G$ and elevate \M20 values. We do not believe that this difference in resolution will impact our results, since \citet{lotz08b} use these boundaries up to $z=1.2$ with little change in their effectiveness, and the change in spatial resolution from $z=1.2$ to $z=2$ is insignificant. Second, the boundaries were determined at a rest-frame central wavelength of $\sim400$\,nm, compared to the $\sim530$\,nm of our sample. The effect of this difference is that our images may be smoother, as they are less affected by extinction and do not sample any emission below the Balmer break. Quantifying the effect of this difference is beyond the scope of this paper, so we simply note it as a caveat. Third, we are assuming that the observed-frame 1.6\,$\mu$m light traces stellar mass, rather than line-emitting gas, in most pixels. Based on previous observations this seems likely \citep[e.g.][]{fors11}, but we cannot exclude the possibility of significant [\ion{O}{3}]5007\AA\ contamination. Fourth, the selection of objects with spectroscopic redshifts may predispose the sample to having more centrally concentrated rest-frame ultraviolet emission (\S\ref{sample}).

Compared to the GNS sample, our sample has markedly different $G$ values; we find that $P_{1,6}^{G} = 0.12\%$, with the $h$DOGs having values higher by $\Delta G\sim 0.17$, on average, indicative of more centrally concentrated light distributions. The \M20 values of our sample are also different to those of the GNS sample; we find that $P_{1,6}^{M_{20}} = 2.2\%$, with the values lower in our sample by $\Delta$\M20$\sim0.35$, indicating that the brightest 20\% of light in our sample shows less variance and is more concentrated into a small number of nuclei. Conversely, the asymmetries of the $h$DOGs are similar to those of the GNS sample, with $P_{1,6}^{A} = 99.5\%$. 

Other studies similar to the GNS exist; while these studies do not tabulate their data, we can still make comparisons. Compared to the $M_{*}\gtrsim10^{10}$\,M$_{\odot}$ systems at $z\lesssim 2$ in \citet{lee13}, our sample have comparable $G$ but slightly more negative \M20 values than their quiescent samples, but higher $G$ and more negative \M20 values than their star-forming samples. Compared to the $M_{*}\gtrsim10^{11}$\,M$_{\odot}$ systems at $z\sim 2$ selected via extremely red, ``IERO'' colors ([$z_{850}$] $-$ [3.6] $>$ 3.25 and [3.6] $<$ 21.5, \citealt{wang12}), our sample have comparable $G$ and slightly more negative \M20 values than their quiescent sample, but higher $G$ and substantially more negative \M20 values than their star forming sample.

Comparisons to the $p$DOGs and $b$DOGs can only be made in the first panel of Figure \ref{fig:gm20a} as these two samples do not have asymmetry measures. Both the $p$DOGs and the $b$DOGs separate from the $h$DOGs in the $G$-\M20 plane, with lower $G$ values and less negative \M20 values, though the $p$DOGs are closer to our sample in \M20 than are the $b$DOGs. We find, for the comparison with the $b$DOGs, $P_{1,2}^{G} = 0.56\%$ and $P_{1,2}^{M20} = 0.23\%$, while for the $p$DOGs we find $P_{1,3}^{G} = 0.07\%$ and $P_{1,3}^{M20} = 9.7\%$. Overall, the $h$DOGs have more concentrated {\itshape and} less asymmetric inner light distributions than do either other class of DOG. 

Finally, we compare our sample to the two SMG samples. The distribution of the A13 SMGs is wider than our sample in all three panels of Figure \ref{fig:gm20a}; our sample lies only within part of the SMG distribution, corresponding to lower than average asymmetries. Conversely, the $G$ and \M20 values for the two samples are (marginally) comparable. We find; $P_{1,5}^{G} = 66.4\%$, $P_{1,5}^{M20} = 54.9\%$, and $P_{1,5}^{A} = 35.1\%$. We see no dependence of AGN fraction for our sample in terms of their position within the A13 SMG distribution. Conversely, the B12 SMGs are offset from our sample, and lie close to the other two DOG samples. We find $P_{1,4}^{G} = 1.1\%$, $P_{1,4}^{M20} = 2.4\%$. Both SMG samples also avoid the bulk of the GNS sample in all three panels.

\begin{figure*} 
\begin{center}
\includegraphics[width=7.5cm,angle=0]{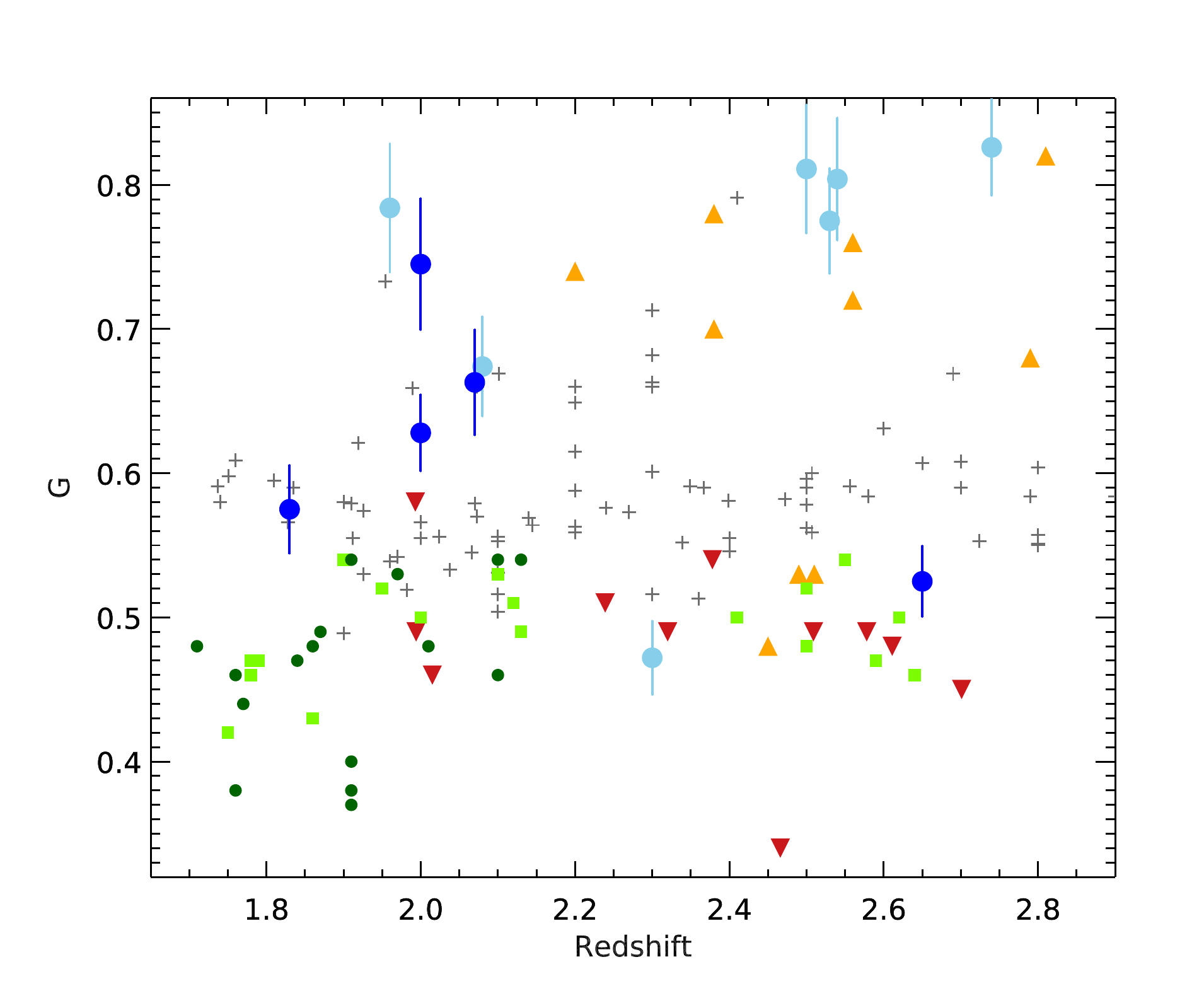} 
\includegraphics[width=7.5cm,angle=0]{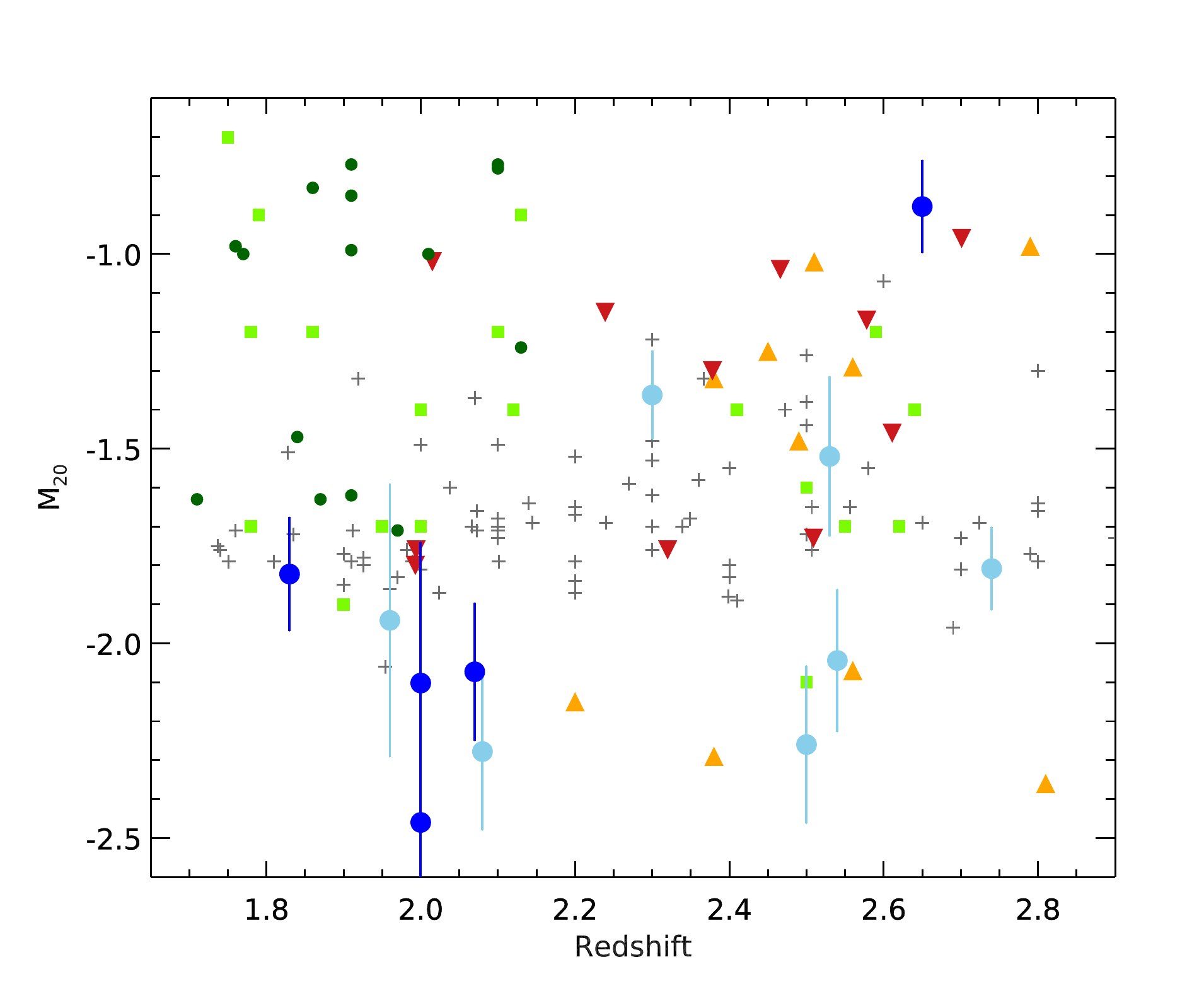} \\
\includegraphics[width=7.5cm,angle=0]{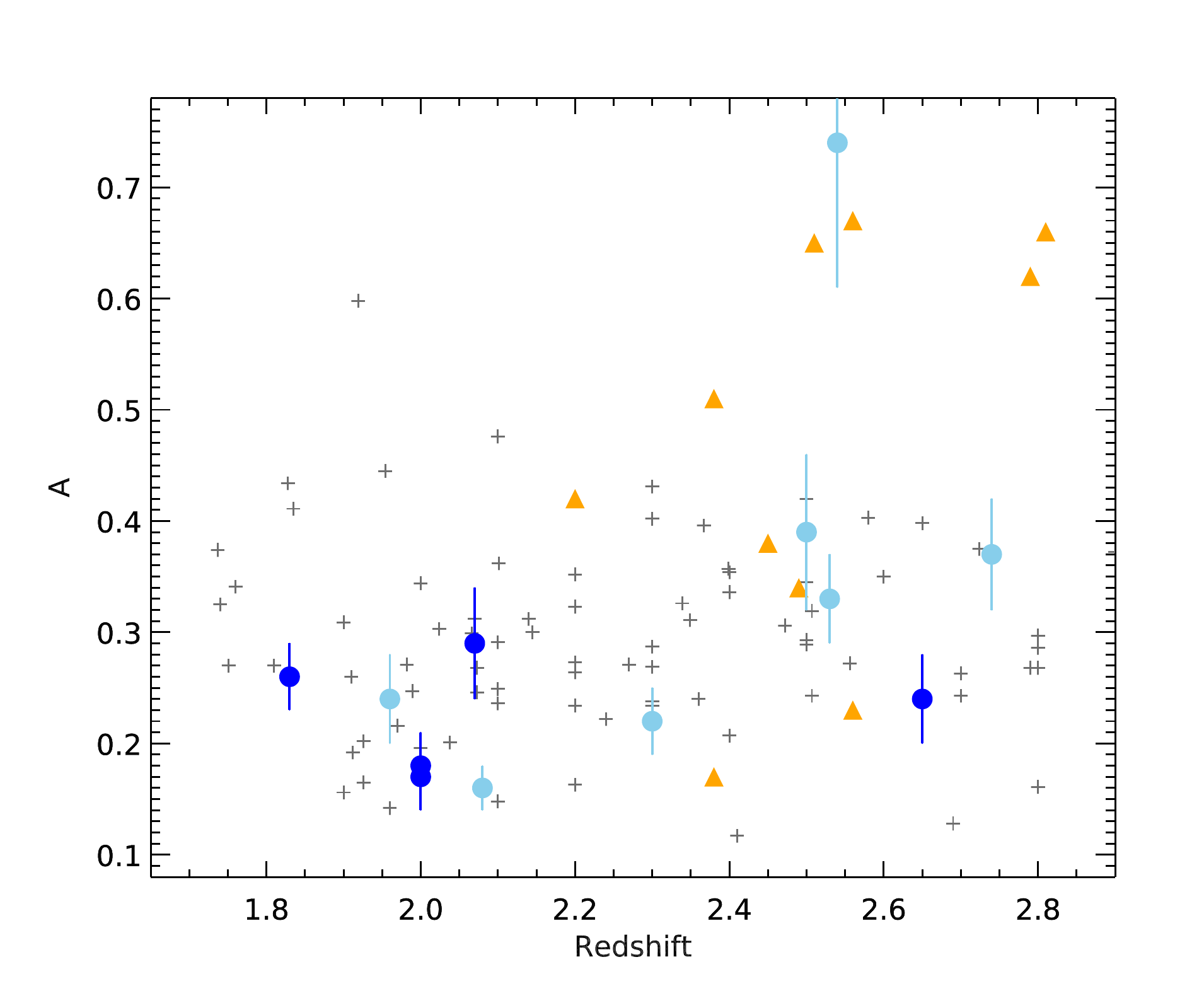} 
\includegraphics[width=7.5cm,angle=0]{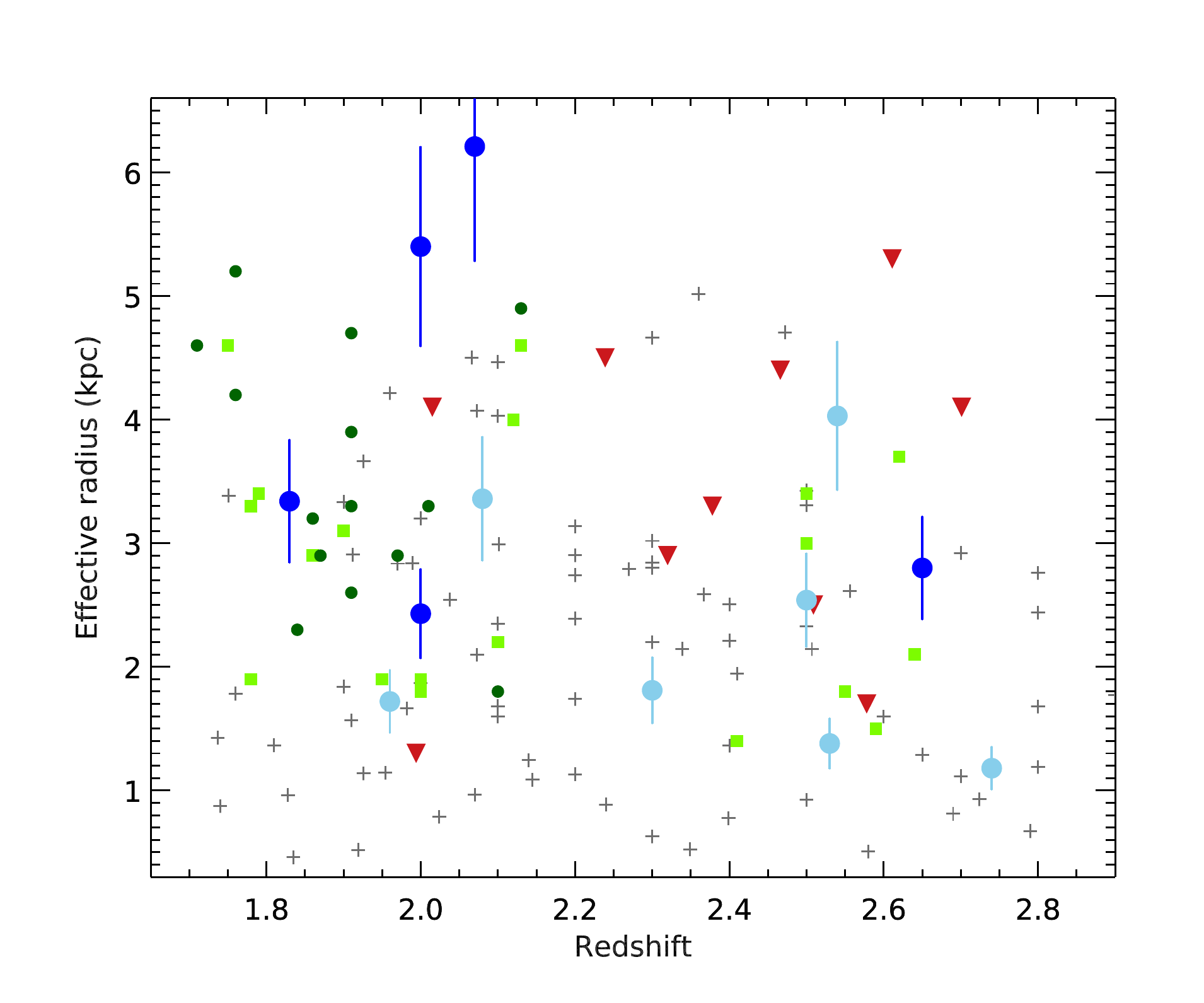} \\ 
\includegraphics[width=7.5cm,angle=0]{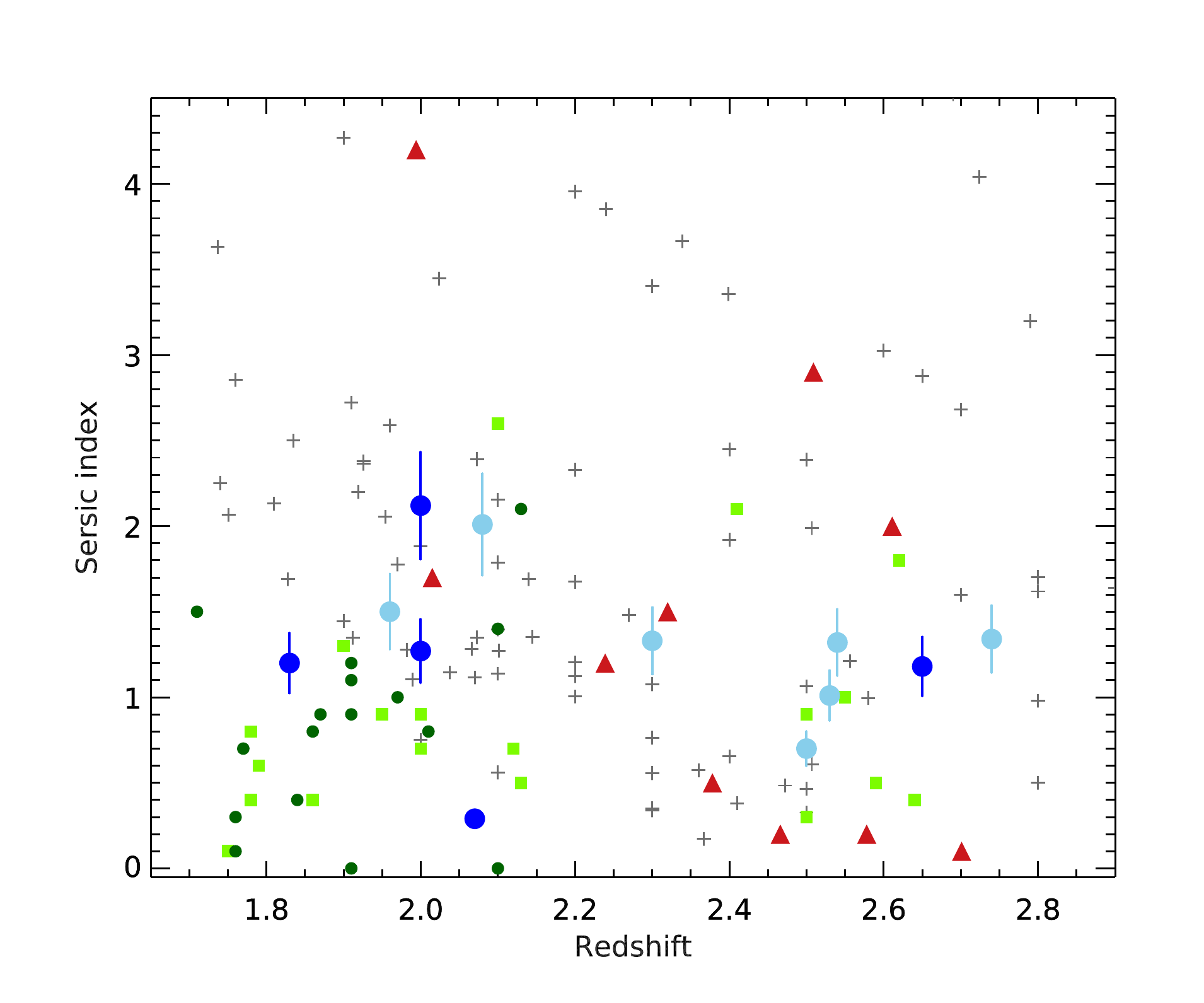} 
\caption{The five morphological parameters of our sample (\S\ref{submorph}), plotted as functions of redshift. See Figure \ref{fig:gm20a} for a key to the points, and \S\ref{compops} for a description of the comparison populations.}
\label{fig:redshiftvsmorph}
\end{center}
\end{figure*}

We next examine the relations between morphological parameters and redshift (Figure \ref{fig:redshiftvsmorph}). No trends are apparent, in any parameter\footnote{We do not consider the effects of surface brightness dimming with redshift here; our sample span $\Delta z=0.91$ so such effects are likely to be small in comparison with other sources of error.}. There are also no clear trends with AGN fraction. Finally, there are no clear trends among any of the comparison populations. This is consistent with the $h/b/p$DOG and SMG selections isolating brief phases in luminous galaxy evolution, and with the physical processes that these selections signpost not changing substantially over $1.8<z<2.7$.

\begin{figure*} 
\begin{center}
\includegraphics[width=19.5cm,angle=0]{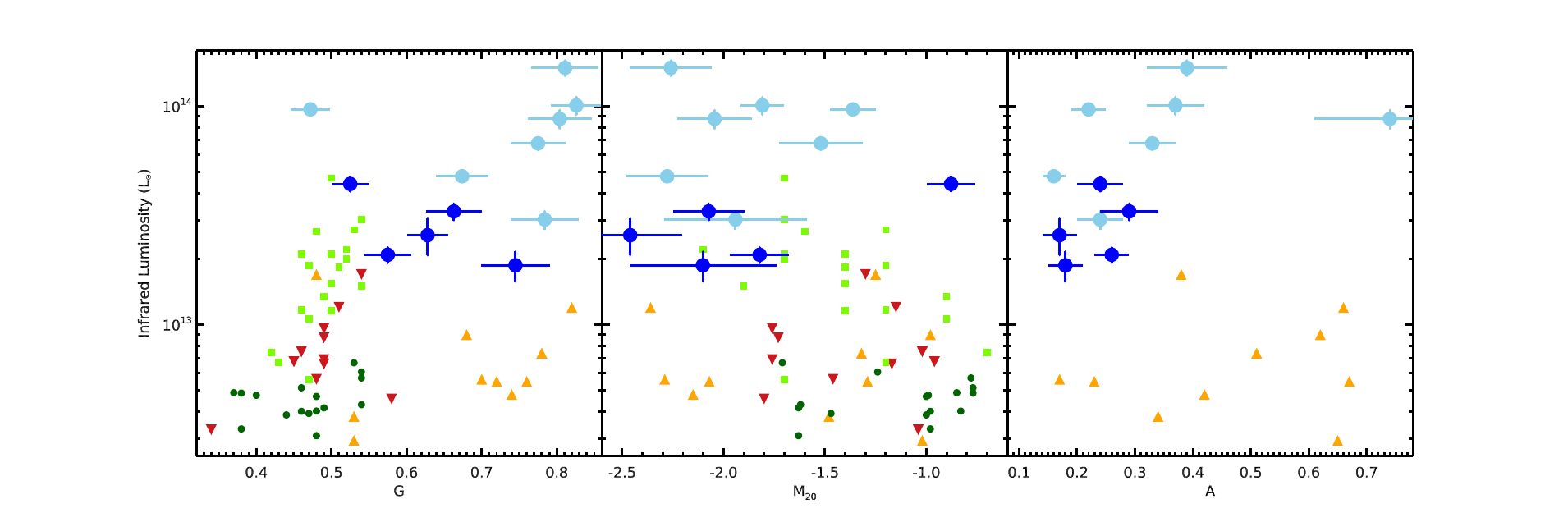}  
\includegraphics[width=13.0cm,angle=0]{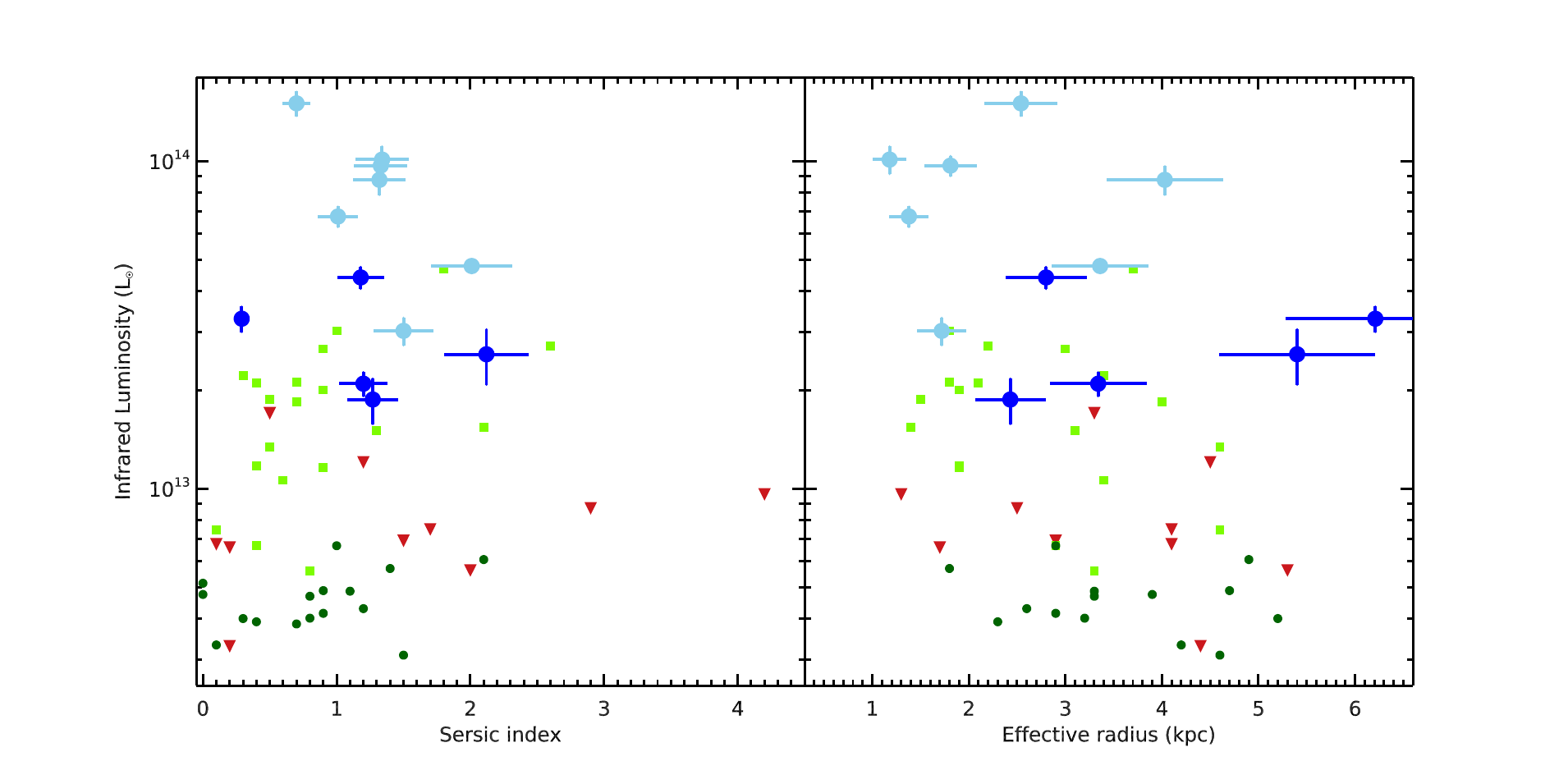} 
\caption{The five morphological parameters of our sample, plotted as a function of infrared luminosity. See Figure \ref{fig:gm20a} for a key to the points, and \S\ref{compops} for a description of the comparison populations..}
\label{fig:redshiftvslir}
\end{center}
\end{figure*}

Next, we search for trends of morphological parameters with infrared luminosity (Figure \ref{fig:redshiftvslir}). No trends are evident in any parameter. We also see no trends in any of the comparison populations. There is, perhaps, a trend between L$_{IR}$ and $G$ if our sample and the $p$DOGs are considered together, but the trend is not strong. Finally, we see no trends of morphological parameters with AGN fractional luminosity, except, perhaps, with asymmetry, where all the objects with $f\mathrm{_{AGN}}>0.95$ have $A<0.35$. As with the lack of trends with redshift, this is consistent with the DOG and SMG selections isolating brief phases in the duty cycle of active galaxies, and/or that infrared luminosity trends do not trace galaxy assembly processes.

\subsection{S{\'e}rsic profiles and effective radii}\label{results2}
Most of our sample have S{\'e}rsic indices close to unity. Two objects have $n\sim2$, and one has $n<0.5$. This is at the lower end of the range of S{\'e}rsics of all galaxies, consistent with them being disk or merger systems. They are not ellipticals or systems with de Vaucoleurs profiles. Smaller, lower luminosity ellipticals can have $n < 4$, with values as low as unity \citep{caon93}, but our sample is unlikely to be small ellipticals with low S{\'e}rsic indices, given that their effective radii are all $>1.15$\,kpc. One classification scheme that uses the S{\'e}rsic index is that of \citet{ravi06}, who propose that mergers have $\langle n \rangle < 0.8$, exponential profile systems have $0.8 < \langle n \rangle < 2.5$, and bulge systems have  $\langle n \rangle > 2.5$. According to this scheme three of our objects are mergers, while the rest are disklike. 

Figure \ref{fig:sersvsall} compares the $G$, \M20, $A$, and $r_{e}$ measurements for our sample to their S{\'e}rsic indices. We see no trends, except that the two objects with $n\simeq2$ have the most negative \M20 and lowest asymmetries, consistent with these two objects being the most dynamically relaxed of the sample. We also see no trends in S{\'e}rsic index with AGN fraction. Compared to the other samples; our sample have somewhat dissimilar S{\'e}rsic indices to the two DOG samples ($P_{1,2}^{n} = 35.4\%$ and $P_{1,3}^{n} = 18.7\%$, see also \citealt{sch12}) but are similar to both the GNS sample ($P_{1,6}^{n} = 99.8\%$) and the B12 SMGs ($P_{1,4}^{n} = 99.7\%$). Compared to the \citet{lee13} sample; the $h$DOGs have comparable S{\'e}rsic indices to their star-forming $M_{*}\gtrsim10^{10}$\,M$_{\odot}$ systems, but lower by $\Delta n\simeq 1$ than their passive systems. The $h$DOG S{\'e}rsic indices are also comparable to those of `main-sequence' star-forming galaxies (with SFRs in the range $20-350$M$_{\odot}$yr$^{-1}$) at $z\sim2$ (\citealt{tacc15}, their $H$-band single component fits, see also \citealt{mori14}). This similarity indirectly suggests that the star formation in $h$DOGs can be located at kpc-scale galactocentric distances. 

Using rest-frame optical spectroscopy, \citealt{wu16} propose that $h$DOGs have 
black hole masses of order $10^{9}$\,M$_{\odot}$ and Eddington ratios close to unity. We compare these measurements to the black hole masses that are {\itshape predicted} for our sample, based on locally observed relations between $n$ and black hole mass. These relations arise because local samples are dynamically relaxed, with a stable bulge. This condition is not satisfied in our sample. Nevertheless, we explore its consequences. Applying the log-normal relation of \citet{gradriv07} yields predicted black hole masses for our sample of order $10^{7}$\,M$_{\odot}$, or two orders of magnitude below the measured average. This is consistent with the idea that the black holes assembled at least a few hundred Myr before the bulges in these systems. 

We compare the effective radii of our sample to the $G$ -- \M20 -- $A$ parameters in Figure \ref{fig:reffvsgm20a}. No trends are apparent. The effective radii of our sample are comparable in distribution to all four of the comparison populations (column 4 of Table \ref{table:probsim}). They are also comparable to other samples of massive, star-forming galaxies at similar redshifts \citep{mori14,tacc15}, and larger than the compact star-forming or quiescent systems at $z\lesssim2$ \citep{dad05,vandok08,vand10,wein11}. The effective radii of our sample are also similar to those of cluster galaxies at $z=1.62$, but have lower S{\'e}rsic indices, on average \citep{papo12}.

\section{Discussion} \label{discon}
The luminosities, colors, power sources and morphologies of our sample are consistent with $h$DOGs signposting a brief but important stage in SMBH mass assembly, during the peak epoch of galaxy assembly \citep{bell12,andr13,bruce14,mori15}. We cannot constrain the stellar masses of our sample, so in the following we assume a stellar mass range of $10^{10-11}$\,M$_{\odot}$ \citep{ass15}. We also note two caveats. First is in the comparisons to the morphological measurements from the literature. While we have matched the samples as closely as possible in redshift and (morphological) rest-frame wavelength, there remain differences in approach to the measurements, such as the treatment of uncertainties, that could introduce systematics between the comparison samples. Second, all the comparison samples have lower infrared luminosities than the $h$DOGs, so luminosity-driven differences are possible. 

We frame the following discussion in terms of three candidate evolutionary scenarios for $h$DOGs:

\begin{itemize}

\item First, that the $h$DOGs are predominantly major ($\lesssim4:1$) mergers, {\itshape and} that they are `exceptional' such systems - e.g. mergers with atypical progenitor properties, or mergers caught in a particular phase, such as late-stage mergers in which star formation is fading. 

\item Second, that $h$DOGs are predominantly mergers, but are drawn at random from the merger population at $z\sim2$. 

\item Third, that $h$DOGs are not preferentially associated with mergers, and are instead drawn from a broad subset of the massive galaxy population at $z\gtrsim2$.

\end{itemize}

The primary evidence that the $h$DOGs are predominantly mergers are their positions in the $G$ -- \M20 -- $A$ plane (Figure \ref{fig:gm20a}), using canonical morphological classification boundaries \citep{con03,lotz04,lotz08a,lotz08b}. In the $G$ -- \M20 plane, nine sources are classified as mergers, one source is classified as early type, one as late-type, and one is ambiguous between all three types. In the $G$ -- $A$ plane, two sources are classified as non-mergers (the early and late type sources from the $G$ -- \M20 plane) while all the rest are either mergers, or close to the mergers boundary. This is consistent with a higher merger fraction than in the massive galaxy population at $z\sim2$. Moreover, our sample have more peaked, more symmetric {\itshape central} light distributions than the $b$DOGs, $p$DOGs, or (most of) the SMGs, which is consistent with the $h$DOGs being more advanced mergers, on average, than any of the $b$DOGs, $p$DOGs, and SMGs. The hint of an anticorrelation between infrared luminosity and AGN fractional luminosity (Figure \ref{fig:lirvszfrac}) could also support this idea, if the peak starburst luminosity occurs before the peak AGN luminosity during a merger. Additional evidence includes: (1) other infrared-luminous samples often have high merger fractions, at both low \citep{petty14,psy16} and high \citep{farrah02,kar12} redshift, with late-stage mergers showing higher obscuration levels than early-stage mergers \citep[e.g.][]{ricc17}, (2) reddened quasars at $z\sim2$, which may be the immediate descendants of the $h$DOGs, have merging hosts in nearly all cases \citep{urr08,gli15,hilb16}, (3) an increased fraction of disturbed or interacting morphologies with increased obscuration has been found for AGN at $z\sim1$ \citep{koc15}, and (4) some simulations suggest that major mergers are the main mechanism for bulge growth at high redshift \citep{fiac15}. 

Overall, this implies that $h$DOGs are predominantly ``exceptional'' late-stage mergers, and that they may be (1) the descendants of some fraction of the (merger-driven) SMGs at $z\sim2.5$, and (2) the antecedents of some fraction of both red quasars at $z\lesssim2$ and evolved massive galaxies at $z<2$. This scenario is consistent with that posited by \citet{fan16} for an overlapping but distinct sample. The rarity of $h$DOGs compared to SMGs and luminous unobscured quasars then implies that $h$DOGs are intrinsically rare - perhaps ones with anomalously high initial gas fractions or gas-rich Mpc-scale environments - and/or that the $h$DOG phase is brief compared to the SMG and unobscured quasar phases.

\begin{figure*} 
\begin{center}
\includegraphics[width=16cm,angle=0]{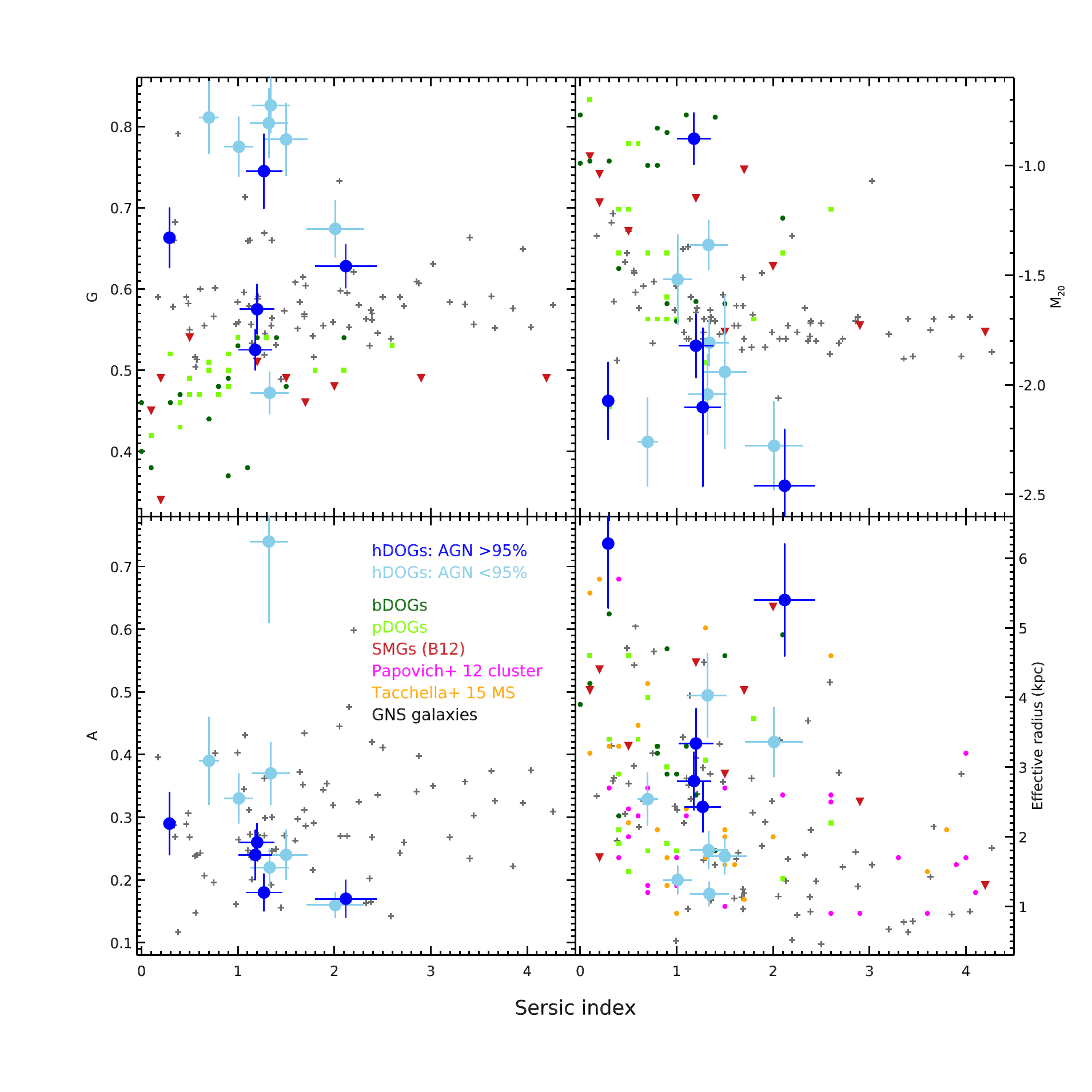} 
\caption{The $G$, \M20, $A$ and $r_{e}$ values for our sample, plotted as functions of S{\'e}rsic index. See \S\ref{compops} \& \S\ref{results2} for a description of the comparison populations.}\label{fig:sersvsall}
\end{center}
\end{figure*}

We do not, however, consider this explanation satisfactory, for four reasons. 

\begin{figure*} 
\begin{center}
\includegraphics[width=20cm,angle=0]{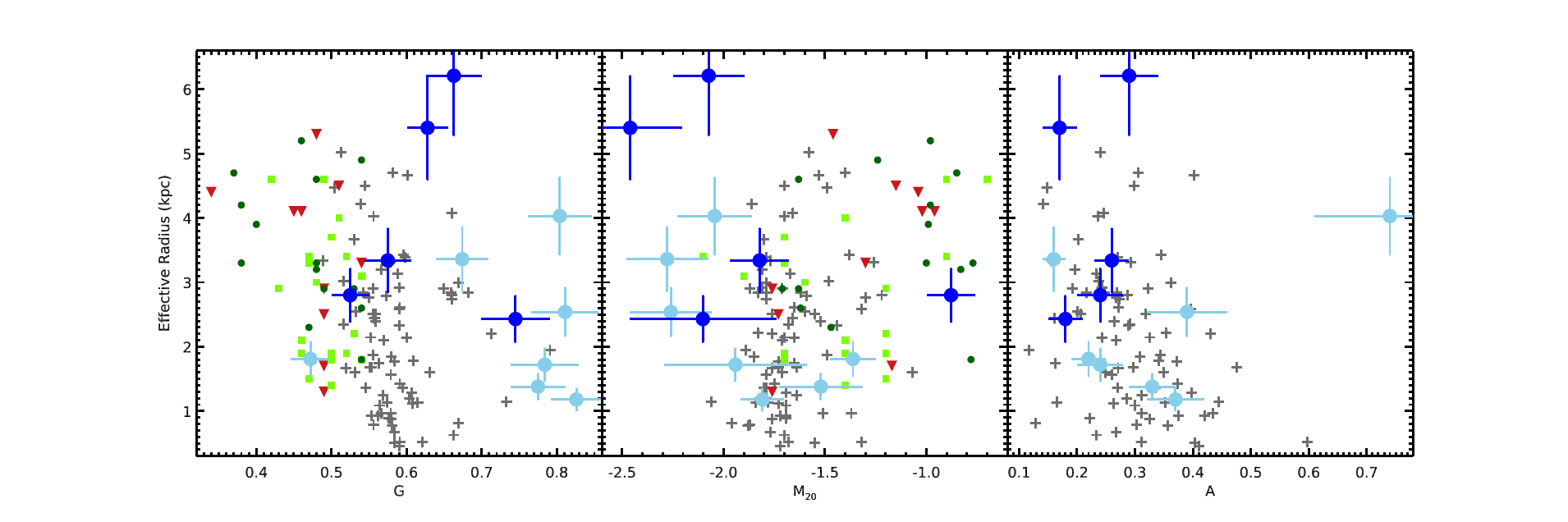} 
\caption{The $G$, \M20, and $A$ values for our sample, plotted as a function of effective radius. See Figure \ref{fig:gm20a} for a key to the points, and \S\ref{compops} for a description of the comparison populations.}
\label{fig:reffvsgm20a}
\end{center}
\end{figure*}

First, the evidence that $h$DOGs are predominantly mergers is less convincing when the asymmetries, effective radii, and S{\'e}rsic indices are considered individually. The $h$DOGs have identical {\itshape total} asymmetries to the GNS sample, but dissimilar asymmetries to the SMGs. Moreover, the effective radii and S{\'e}rsic indices of the $h$DOGs are much closer in distribution to the GNS galaxies than to the other three comparison samples. This is consistent with the {\itshape galaxy-wide} gravitational potentials of the $h$DOGs most closely resembling those of the GNS sample. Furthermore, if the $h$DOGs are late-stage mergers, then we might expect to see elliptical profiles in those systems with relaxed profiles, since massive, quenched galaxies are mainly bulge dominated at $z \lesssim 2$ \citep{bell12,wuyts12,lang14,bruce14,manc15,huer16}. However, we do not observe the $\langle n \rangle \gtrsim 4$ S{\'e}rsic indices of bulges in any of our sample. Finally, in all panels of Figures \ref{fig:gm20a}, \ref{fig:sersvsall}, \ref{fig:reffvsgm20a} the $h$DOGs appear {\itshape uniquely} distributed - while they may resemble other populations we consider in a single morphological parameter, they are dissimilar to all of them in any combination of morphological parameters. 

Second, the infrared luminosities, AGN fractional luminosities, and effective radii of our sample show no trends with any morphological parameter. This argues against any morphological parameter {\itshape coherently} tracing an advancing merger\footnote{One way to alleviate this argument is that $h$DOGs are a brief phase in the lifetime of a merger, {\itshape and} are triggered at random points in the merger, so as to obtain the wide spread in $G$, \M20, and $A$ values that we see.}. We also see no trends in S{\'e}rsic index with any other parameter, arguing against mergers driving the formation of bulges.

Third, the use of the $G$-\M20 plane to classify mergers {\itshape at high redshift} is problematic. High redshift galaxies that contain luminous, off-center `clumps' can resemble interacting systems in $G$-\M20 space, even though they are not interacting \citep[e.g.][]{fors06,thom16}. 

Fourth, recent simulations propose that asymmetry alone is a better discriminant of major mergers than $G$ or \M20, and objects with $A>0.8$ are mergers, rather than the canonical cut of $A>0.35$ \citep{thom16}. Using this higher value, none of our sample are mergers. Using the canonical value \citep{con03}, only three of our sample are mergers.

A plausible alternate scenario is as follows. In a massive, star-forming galaxy at $z\sim2$, a luminous AGN raises $G$ and lowers \M20 relative to the massive star-forming galaxy population, but does not appreciably affect any of $A$, $r_{e}$, or $n$. This scenario is compatible with the work of \citet{pierce10}, who show that an AGN can elevate $G$ by up to 0.2, and reduce \M20 by up to 1.0, while leaving $A$ and $n$ unchanged (see also \citealt{simurr08}). Such a briefly luminous, or `flickering', AGN placed pseudo-randomly within the massive galaxy population at $z\sim2$ would also give no correlation between morphologies and AGN fractional luminosities, as we observe\footnote{The ubiquity of mergers in luminous, obscured AGN at $z\lesssim0.4$ \citep[e.g.][]{farrah01} may seem inconsistent with this, but this could arise due to e.g. less rapid dark matter density contrast evolution, or a lower free gas fraction, thus requiring specific mechanisms to trigger luminous activity.}. 

This scenario implies that the nuclear $H$-band light arises at least in part from the AGN. Even if this is not the case though, this scenario is still plausible. It has been suggested that quenching correlates most strongly with {\itshape central} ($\lesssim1$kpc) stellar mass surface density, rather than total stellar mass or S{\'e}rsic index \citep{cheung12}. Thus, if the $G$-\M20 positions of the $h$DOGs do not arise from AGN light, but rather from the {\itshape earliest} formation stages of a bulge, then this could be evidence for such a transition, and could link $h$DOGs to the small fraction of bulge dominated star-forming galaxies at $z\sim2$ \citep{bruce14a}, as well as SMGs \citep{simp15}. If so though, it still does not argue for a preferential association with mergers. 

This scenario is consistent with studies that propose that mergers are not the main mode for massive galaxy assembly at $z\gtrsim2$ \citep{wang12,loft17}, and that mergers are not the only trigger for AGN, even at high luminosities \citep{vill14,vill16}. For example, it has been suggested that there exist two `channels' for bulge assembly in massive galaxies; (1) a rapid channel at $z>3$, since some bulges are already in place with high S{\'e}rsic indices at $z=2.5$, and (2) a gradual channel, transitioning from clumpy disks to bulge+disk systems, at $1<z<3$ \citep{huer15}. The morphologies of the $h$DOGs are consistent with this second track. Moreover, \citet{brennan15} propose that models which include a channel for bulge growth via disc instabilities agree better with observations than models in which bulges can grow only through mergers. Furthermore, SMGs may not be exclusively mergers, but instead include `extreme' examples of normal star-forming galaxies \citep{targ13}. Finally, \citet{sch12} propose that only a small fraction of DOGs are mergers, with most being disklike\footnote{Though they do suggest that there may be a higher fraction of mergers at high luminosities.}. Finally, this scenario is consistent with that suggested by \citet{cima13} for lower luminosity AGN at similar redshifts to our sample. Thus, even if we posit that $h$DOGs lie on the same evolutionary path as one or more of SMGs, $p$DOGs, and $b$DOGs, it is {\itshape still} not necessary to invoke a {\itshape preference} for mergers. The $h$DOGs can still be the ancestors of low redshift massive ellipticals, since there is evidence for multi-stage formation in this population \citep{petty13}. There is also no tension with the properties of extremely red quasars (ERQs) at $z\gtrsim1$; the ERQs show evidence for powerful outflows and some fraction of the ERQs have similar colors to the DOGs \citep{ross15,hama17}, but the ERQs with merging hosts could be the fraction of obscured AGN that are triggered by mergers. 

Overall, with the caveat that our sample is small, we find the second scenario more convincing. We thus conclude that $h$DOGs are, as a class, a brief and/or rare stage in massive galaxy assembly at $z>2$, but that they are most likely a phase in which a luminous AGN turns on in a massive, star-forming galaxy. Depending on the origin of the nuclear $H$-band light, this phase may also include the earliest formation stages of a bulge. We do not however find that $h$DOGs are {\itshape preferentially} mergers. This is in contrast to the conclusions drawn by \citet{fan16}. A significant merger fraction is however still plausible, since the $1<z<3$ epoch exhibits a higher merger fraction than lower redshifts \citep{lotz11}, and since star formation may be triggered by mergers in up to 27\% of massive galaxies at $z=2$ \citep{kav13}.

Finally, we speculate on a link between $h$DOGs and compact star-forming galaxies. \citet{barr14} find that nearly half of all compact $>10^{10}$\,M$_{_{\odot}}$ star forming galaxies at similar redshifts to our sample host an AGN with $L_{X} \gtrsim 10^{43}$\, ergs s$^{-1}$. This fraction of sources that host an AGN is lower than found in the $h$DOGs, but higher than in main-sequence star-forming galaxies at $z\sim2$. Moreover, the S{\'e}rsic indices of these compact galaxies are similar to our sample. It is thus plausible that our sample are the immediate antecedents to the compact star forming and quiescent galaxies, that peak in number at $z\sim2$. This idea is also consistent with the $h$DOGs not being preferentially associated with mergers, since the dissipational events that lead to compaction can be either gas-rich mergers, or disk instabilities triggered by gas accretion \citep{hop08,dek09,wellons15}. Moreover, \citealt{willi14} find that the morphologies of compact Early-type galaxies at $z=1.6$ are inconsistent with a major merger origin. If this is the case then it may mean that AGN are important in compactifying galaxies at high redshift (\citealt{fan08,fan10,chang17}, though see also \citealt{Lilly16}), which would allow for compaction before a galaxy loses the bulk of its gas and dust, as found by \citet{barr14}. However, since the Barro et al sample is approximately four orders of magnitude more numerous on the sky than the $h$DOGs, and their AGN are at least three orders of magnitude less luminous than those in the $h$DOGs, our sample could only represent a tiny fraction of the antecedents of compact star forming galaxies. 

\section{Conclusions} \label{conclude}
We have presented HST WFC3 F160W imaging, and fits to the infrared spectral energy distributions, for twelve extremely luminous, obscured AGN at $1.8<z<2.7$. Our conclusions are:

1 - The infrared luminosities of our sample lie in the range $2-15\times10^{13}$L$_{\odot}$, making them among the most luminous objects in the Universe at $z\sim2$. In all cases the infrared colors and SED fits are consistent with the infrared emission arising at least in most part from obscured AGN activity. Star formation rates of up to several hundred Solar masses per year are however still plausible in most objects. The AGN fractional luminosities are higher than is seen in infrared-luminous galaxies selected either via their submillimeter emission, or via an $R--24\mu$m color cut. 

2 - The morphologies of our sample, sampling rest-frame $B$-band light, appear mostly clumpy and irregular. Their Gini coefficients span $0.47 < G < 0.81$, their \M20 parameters nearly all lie in the range $-2.5 <$ \M20 $< -1.3$, and their asymmetries are nearly all in the range $0.16 < A < 0.39$. These numbers are consistent with our sample being moderately asymmetric, but with markedly concentrated and regular light profiles in their central regions. The effective radii of our sample span 1-6\,kpc, making them comparable in size to near-infrared selected massive galaxies at $z\sim2$. We see no dependence of AGN luminosity, or AGN fractional luminosity, on any of $G$, \M20, $A$, $n$, or $r_{e}$. 

3 - The S{\'e}rsic indices of our sample span $0.25 < n < 2.15$. Combined with spectroscopic black hole mass measurements from \citet{wu16}, this is consistent with the mass assembly of the central black holes in our sample leading the mass assembly of any bulge component. 

4 - Based on canonical classification boundaries in the $G$-\M20-$A$ plane, most of our sample are mergers. We do not, however, believe that this is the most plausible scenario. Our sample have comparable asymmetries, effective radii and S{\'e}rsic indices to main-sequence star-forming galaxies at $z\sim2$. Together with the lack of trends between morphological parameters and AGN properties, this implies that our sample are drawn from the massive, star-forming galaxies at $z\sim2$ which harbor a briefly luminous, ``flickering'' AGN, and potentially the earliest formation stages of a bulge in an inside-out manner. While a significant merger fraction is still plausible, we find no need to invoke a {\itshape preferential} link with mergers. The divergent $G$ and \M20 values of our sample compared to massive, star-forming galaxies at $z\sim2$ can also be explained via observed-frame near-infrared emission in the central regions either from the AGN, or from a nascent bulge component. 

5 - We speculate that our sample may represent a small fraction of the immediate antecedents to compact star forming galaxies at $z\lesssim2$. The compact galaxies have a lower AGN fraction than our sample, but higher than observed in extended star-forming galaxies at $z>2$. Moreover, they have comparable S{\'e}rsic indices to our sample. If the AGN in our sample are responsible for triggering compaction, and potentially also the formation of a bulge, then this would help explain why compact star forming galaxies can still harbor large masses of gas and dust.

\acknowledgments
The authors thank the referee for a very helpful report, and acknowledge support from NASA Grant HST-GO-12585. This publication makes use of data products from the \textit{Wide-field Infrared Survey Explorer}, which is a joint project of the University of California, Los Angeles, and the Jet Propulsion Laboratory/California Institute of Technology, and NEOWISE, which is a project of the Jet Propulsion Laboratory/California Institute of Technology. WISE and NEOWISE are funded by the National Aeronautics and Space Administration. This research has made use of the NASA/ IPAC Infrared Science Archive, which is operated by the Jet Propulsion Laboratory, California Institute of Technology, under contract with the National Aeronautics and Space Administration. The National Radio Astronomy Observatory is a facility of the National Science Foundation operated under cooperative agreement by Associated Universities, Inc. RJA was supported by FONDECYT grant number 1151408.

\end{document}